\documentclass[a4paper,12pt]{article} 
\usepackage{color}
\usepackage{epsfig}
\usepackage{amsfonts}
\usepackage{amsmath}
\usepackage{amssymb}
\usepackage{rotating}
\usepackage{multirow}
\usepackage{mathtools}
\usepackage{chngpage}
\usepackage{setspace}
\usepackage{array}
\usepackage{caption}
\usepackage{url}
\DeclareMathOperator{\se}{se}
\DeclareMathOperator{\cov}{cov}

\DeclareMathOperator{\sign}{sign}
\DeclareMathOperator{\dopearl}{do}
\newcommand\independent{\protect\mathpalette{\protect\independenT}{\perp}}
\def\independenT#1#2{\mathrel{\rlap{$#1#2$}\mkern2mu{#1#2}}}
\begin{document}
\title{Assessing the effectiveness of robust instrumental variable methods using multiple candidate instruments with application to Mendelian randomization}
\author{Stephen Burgess \textsuperscript{1,} \thanks{Address: Department of Public Health and Primary Care, Strangeways Research Laboratory, 2 Worts Causeway, Cambridge, CB1 8RN, UK. Telephone: +44 1223 748651. Correspondence to: sb452@medschl.cam.ac.uk.} \and Jack Bowden \textsuperscript{2} \and Frank Dudbridge \textsuperscript{1,3} \and Simon G.\ Thompson \textsuperscript{1} \and \\
\textsuperscript{1} Department of Public Health and Primary Care, \\ University of Cambridge, Cambridge, UK. \\ \\
\textsuperscript{2} Integrative Epidemiology Unit, \\ University of Bristol, Bristol, UK. \\ \\
\textsuperscript{3} Department of Non-communicable Disease Epidemiology, \\ London School of Hygiene and Tropical Medicine, London, UK.\\}
\maketitle

\noindent \noindent \textbf{Short title:} Robust instrumental variable methods applied to Mendelian randomization.

\clearpage

\setstretch{1.4}
\subsection*{Abstract}
Mendelian randomization is the use of genetic variants to make causal inferences from observational data. The field is currently undergoing a revolution fuelled by increasing numbers of genetic variants demonstrated to be associated with exposures in genome-wide association studies, and the public availability of summarized data on genetic associations with exposures and outcomes from large consortia. A Mendelian randomization analysis with many genetic variants can be performed relatively simply using summarized data. However, a causal interpretation is only assured if each genetic variant satisfies the assumptions of an instrumental variable. To provide some protection against failure of these assumptions, robust methods for instrumental variable analysis have been proposed. Here, we develop three extensions to instrumental variable methods using: i) robust regression, ii) the penalization of weights from candidate instruments with heterogeneous causal estimates, and iii) L1 penalization. Results from a wide variety of robust methods, including the recently-proposed MR-Egger and median-based methods, are compared in an extensive simulation study. We demonstrate that two methods, robust regression in an inverse-variance weighted method and a simple median of the causal estimates from the individual variants, have considerably improved Type 1 error rates compared with conventional methods in a wide variety of scenarios when up to 30\% of the genetic variants are invalid instruments. While the MR-Egger method gives unbiased estimates when its assumptions are satisfied, these estimates are less efficient than those from other methods and are highly sensitive to violations of the assumptions. Methods that make different assumptions should be used routinely to assess the robustness of findings from applied Mendelian randomization investigations with multiple genetic variants.

\textbf{Keywords:} Mendelian randomization, instrumental variables, robust methods, summarized data, aggregated data, MR-Egger.

\clearpage

\setstretch{1}
\section{Introduction}
\begin{figure}[htb]
\begin{center}
\includegraphics[height=4.5cm, clip=true, angle=270]{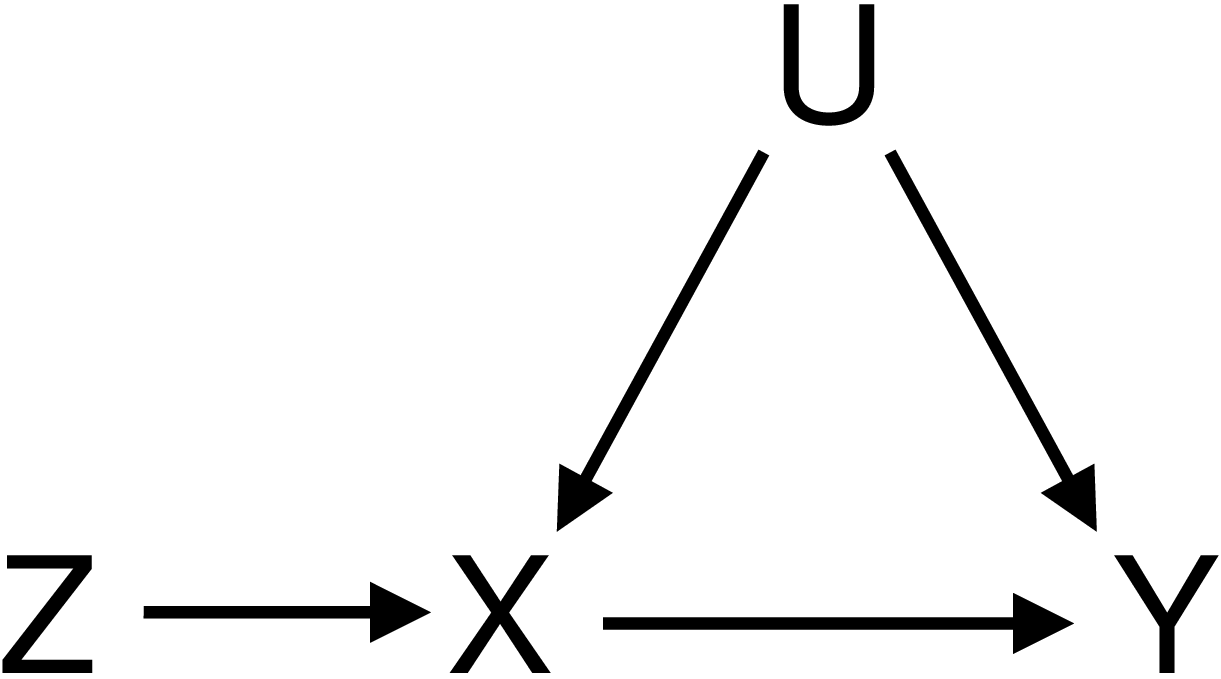}
\end{center}
\caption{Directed acyclic graph of graphical instrumental variable assumptions.} \label{dag}
\end{figure}

An instrumental variable (denoted $Z$) is a randomized or quasi-randomized variable that can be used to estimate the causal effect of an exposure (or risk factor, denoted $X$) on an outcome ($Y$) in the presence of arbitrary unmeasured confounding \cite{greenland2000}. An instrumental variable satisfies three assumptions:
\begin{itemize}
\item[i)]   association with the exposure: $Z \not\independent X$;
\item[ii)]  independence from confounders (denoted $U$) of the exposure--outcome association: $Z \independent U$;
\item[iii)] independence from the outcome conditional on the exposure and confounders: $Z \independent Y | X, U$.
\end{itemize}
The assumptions imply that an instrumental variable cannot have a direct effect on the outcome, but instead any effect is mediated via the exposure (this is known as the exclusion restriction assumption) \cite{clarke2012}. A directed acyclic graph illustrating these assumptions is given as Figure~\ref{dag}.

\begin{center}
[Figure~\ref{dag} should appear about here.]
\end{center}

The instrumental variable assumptions are restrictive and often unrealistic in practice. One way of assessing whether these assumptions are satisfied is to compare the causal estimates from several proposed instrumental variables \cite{small2007}. If the instrumental variable assumptions are satisfied, and under additional parametric assumptions (sufficient conditions are linearity of the instrumental variable--exposure, instrumental variable--outcome and exposure--outcome relationships with no effect heterogeneity), the same causal parameter is estimated by each of the instrumental variables \cite{didelez2007}. In this paper, we use the term `candidate instrument' to describe a variable that is associated with the exposure and hypothesized to satisfy the other instrumental variable assumptions, without prejudicing whether those assumptions are satisfied or not.

A context in which there are often multiple candidate instruments that may plausibly satisfy the instrumental variable assumptions is Mendelian randomization \cite{daveysmith2003, burgess2015book}, the use of genetic variants as instrumental variables \cite{didelez2007, lawlor2007}. For complex (i.e.\ polygenic and multifactorial) exposures, such as body mass index \cite{speliotes2010} or blood pressure \cite{ehret2011}, many associated genetic variants have been discovered in genome-wide association studies. A recent development in Mendelian randomization is the availability of summarized data \cite{burgess2014twosample}. These data comprise the associations (beta-coefficients and standard errors) of genetic variants with the exposure and with the outcome estimated from univariable regression models. Such associations estimated in large sample sizes have been made publicly available for download by many consortia; examples include associations with glycaemic traits from the Meta-Analyses of Glucose and Insulin-related traits Consortium \cite{scott2012} and with coronary artery disease from the CARDIoGRAMplusC4D consortium \cite{nikpey2015}. Instrumental variable methods using summarized data have been recently developed, and include an inverse-variance weighted method \cite{burgess2013genepi}, and two robust methods: a median-based method \cite{bowden2015median} and MR-Egger \cite{bowden2015}; a robust method is defined here as one that can provide reasonable estimates under weaker assumptions than a conventional approach that assumes all candidate instruments are valid.

In this paper, we consider robust methods for causal inference using multiple instrumental variables, focusing on those that can be implemented using summarized data for uncorrelated candidate instruments. This is typical of an applied Mendelian randomization investigation; when the instrumental variable assumptions are in doubt, it is common to include one genetic variant from each gene region in the analysis. These variants will typically be independently distributed as predicted by Mendel's laws due to their physical separation. Although these robust methods have good theoretical properties, there are several issues particularly with the MR-Egger method, such as low power in realistic scenarios \cite{bowden2015median}, and the influence of outlying variants \cite{burgess2016egger}.

In Section~\ref{sec:methods}, we introduce three extensions to existing instrumental variable methods: the use of robust regression, penalization of weights, and L1 penalization. In Section~\ref{sec:simul}, we perform a simulation analysis to compare estimates from various robust methods with respect to bias and coverage properties when some of the candidate instruments do not satisfy the instrumental variable assumptions. Parameters in the simulation analysis are chosen to reflect a typical Mendelian randomization investigation. In Section~\ref{sec:appliedex}, we show how these methods perform in an applied analysis of the causal effect of body mass index on the risk of schizophrenia. We conclude by discussing the results of this paper, and the potential for future developments (Section~\ref{sec:discuss}).

\section{Methods}
\label{sec:methods}
Existing robust methods for causal inference using instrumental variables have taken two different approaches \cite{burgess2016robust}. The first approach (which includes the median-based method \cite{bowden2015median}) is to assume that some, but not all, of the candidate instruments satisfy the instrumental variable assumptions. The second approach (which includes the MR-Egger method \cite{bowden2015}) allows all of the candidate instruments not to be valid instrumental variables, but assumes that the set of variants satisfies a weaker assumption. Software code for implementing all the methods used in this paper is provided in Web Appendix A.1.

\subsection{Parametric assumptions}
We assume throughout that the parametric assumptions of linearity with no effect heterogeneity hold for the causal exposure--outcome relationship, and for the instrumental variable--exposure and instrumental variable--outcome associations for all valid instruments $Z_j$, $j = 1, \ldots, J$:
\begin{align}
\mathbb{E}(X | Z_j=z)         &= \beta_{X0j} + \beta_{Xj} \; z \label{eq:assumpts} \\
\mathbb{E}(Y | Z_j=z)         &= \beta_{Y0j} + \beta_{Yj} \; z \quad \mbox{ for } j = 1, \ldots, J \notag \\
\mathbb{E}(Y | \dopearl(X=x)) &= \theta_{0}  + \theta \; x  \notag
\end{align}
where $\dopearl(X=x)$ is the do-operator of Pearl meaning that the value of the exposure is set to $x$ by intervention \cite{pearl2000}, and the causal effect parameter $\theta = \dfrac{\beta_{Yj}}{\beta_{Xj}}$ for all valid instruments \cite{didelez2007}. Issues relating to the plausibility of these parametric assumptions are left to the discussion.

\begin{figure}[htb]
\begin{center}
\includegraphics[height=2.5cm, clip=true]{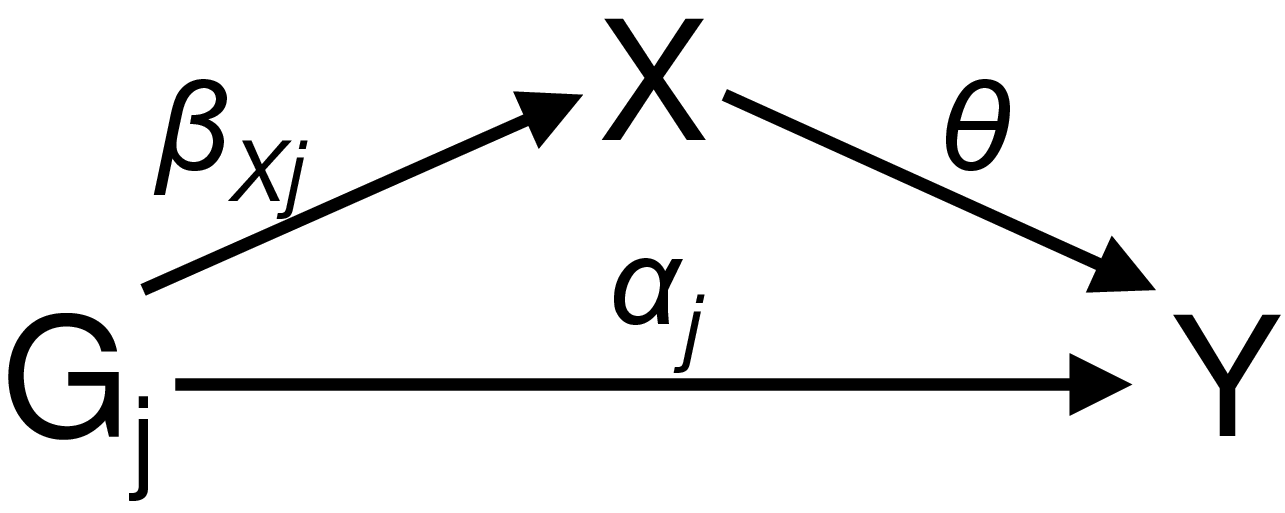}
\end{center}
\caption{Decomposition of association with the outcome $Y$ for genetic variant $G_j$ into indirect (causal) effect via the exposure $X$ and direct (pleiotropic) effect (see equation~\ref{eq:decompose}).} \label{decomp}
\end{figure}

Under these linearity assumptions, the association of a candidate instrument with the outcome $\beta_{Yj}$ decomposes into a direct effect $\alpha_j$ and an indirect effect that corresponds to the causal effect of the exposure on the outcome ($\theta$) multiplied by the association of the candidate instrument with the exposure ($\beta_{Xj}$) \cite{kang2014}:
\begin{equation}
\beta_{Yj} = \alpha_j + \theta \; \beta_{Xj}. \label{eq:decompose}
\end{equation}
The term `pleiotropy' refers to a genetic variant having associations with more than one risk factor on different causal pathways \cite{lawlor2007}. A pleiotropic genetic variant is typically not a valid instrumental variable. In this decomposition, genetic variant $j$ is pleiotropic if $\alpha_j \neq 0$, and $\alpha_j$ is referred to as the pleiotropic effect. The decomposition is illustrated in Figure~\ref{decomp} (the genetic variant is denoted $G_j$ rather than $Z_j$ to emphasize that it is not a valid instrument if $\alpha_j \neq 0$).

The outcome is assumed to be continuous. If the outcome is binary, then the methods can proceed using genetic associations from logistic regression analyses (log odds ratios), provided that the linear assumptions above hold for the logit-transformed probability of the outcome. There are some technical issues with the interpretation of the causal estimate with a binary outcome and a logistic-linear model \cite{burgess2012noncollapse}. However, instrumental variable estimates are typically unbiased under the null in this setting, and Type 1 error rates are not generally inflated (with the notable exception of those from the two-stage residual inclusion method \cite{vansteelandt2010}).

\begin{center}
[Figure~\ref{decomp} should appear about here.]
\end{center}

\subsection{MR-Egger method and the InSIDE assumption}
The MR-Egger method is performed by weighted linear regression of the associations of the candidate instruments with the outcome ($\hat{\beta}_{Yj}$) on the associations of the candidate instruments with the exposure ($\hat{\beta}_{Xj}$), using the inverse of the variances of the associations with the outcome ($\se(\hat{\beta}_{Yj})^{-2}$) as weights:
\begin{equation}
\hat{\beta}_{Yj} = \theta_{0} + \theta_{1} \; \hat{\beta}_{Xj} + \epsilon_{j} \mbox{, weights = } \se(\hat{\beta}_{Yj})^{-2}. \label{eq:egger}
\end{equation}

If the intercept term in this regression model is estimated, then the slope coefficient is the MR-Egger estimate. To ensure that the MR-Egger estimate is invariant to the original arbitrary choice of the coding allele (effect allele) for each genetic variant, we first orientate all the associations so that the $\hat{\beta}_{Xj}$ estimates are all positive \cite{burgess2016egger}. If the intercept term is set to zero, then the slope coefficient is the inverse-variance weighted (IVW) estimate. The IVW estimate can also be obtained by inverse-variance weighted meta-analysis of the ratio estimates $\hat{\theta}_j$ using standard errors $\dfrac{\se(\hat{\beta}_{Yj})}{\hat{\beta}_{Xj}}$. Simulations have shown that this simple choice of standard error expression gives reasonable inferences in realistic settings \cite{burgess2013genepi}. Additionally, the IVW estimate using this expression is the same as the two-stage least squares estimate that can be calculated using individual-level data \cite{burgess2016ivw} (assuming, as throughout, that the candidate instruments are uncorrelated).

The ratio estimate based on candidate instrument $j$ is $\dfrac{\hat{\beta}_{Yj}}{\hat{\beta}_{Xj}}$. This is a consistent estimate of the causal effect if $\alpha_j = 0$. As a weighted mean of the ratio estimates, the IVW estimate is consistent if $\alpha_j = 0$ for all $j$; that is, if none of the candidate instruments are pleiotropic. The MR-Egger method gives consistent estimates of the causal effect under the condition that the pleiotropic effects of the candidate instruments $\alpha_j$ are uncorrelated with the associations of the candidate instruments with the exposure $\beta_{Xj}$ \cite{bowden2015}. This is referred to as the InSIDE assumption (INstrument Strength Independent of Direct Effect). Specifically, we require the weighted covariance $\cov_w(\boldsymbol\alpha, \boldsymbol\beta_{X})$ to be zero:
\begin{equation}
\cov_w(\boldsymbol\alpha, \boldsymbol\beta_{X}) = \frac{\sum_j (\alpha_j - \bar{\alpha}_w) (\beta_{Xj} - \bar{\beta}_{Xw}) \se(\hat{\beta}_{Yj})^{-2}}{\sum_j \se(\hat{\beta}_{Yj})^{-2}} = 0
\end{equation}
where $\bar{\beta}_{Xw}$ is the weighted mean of the $\beta_{Xj}$, $\bar{\alpha}_w$ is the weighted mean of the $\alpha_j$, and bold symbols represent vectors across the candidate instruments.

If the intercept term in the MR-Egger analysis differs from zero, then either the InSIDE assumption is violated, or the average pleiotropic effect differs from zero (referred to as directional pleiotropy). In either case, the instrumental variable assumptions are violated for at least one of the candidate instruments, and the IVW estimate will not be consistent. However, provided that the InSIDE assumption holds, the MR-Egger estimate will still be consistent for the causal effect even in the case of directional pleiotropy. The statistical test of whether the intercept term in MR-Egger differs from zero is referred to as the MR-Egger intercept test.


\subsection{Motivation: robustness to heterogeneity in causal estimates}
The MR-Egger regression has a 100\% breakdown point in the sense that all of the candidate instruments can violate the instrumental variable assumptions by having direct effects on the outcome ($\alpha_j \neq 0$), provided that the InSIDE assumption is satisfied. However, it relies on the InSIDE assumption being satisfied for the complete set of candidate instruments. In contrast, in the simple median method (calculated as the median of the ratio estimates from each candidate instrument individually), up to 50\% of candidate instruments can violate the instrumental variable assumptions arbitrarily \cite{bowden2015median}. It would be worthwhile to develop a method that gives robust estimates if either of these two assumptions holds. We propose three novel extensions to existing instrumental variable methods to explore this possibility: 1) robust regression, 2) penalization of weights, and 3) L1 penalization. All approaches downweight the contribution to the causal estimate of candidate instruments with heterogeneous ratio estimates.


\subsection{Robust regression}
Several methods have been proposed for performing robust regression (that is, regression with greater than a 0\% breakdown point) \cite{huber2011}. Here, we use an MM-estimation approach as described by Koller and Stahel \cite{koller2011}. Each of the letter Ms refers to a ``maximum likelihood type'' maximization step. Briefly, the method proceeds by finding a robust S-estimate (``scale-type estimate'') that minimizes an M-estimate of the scale of the residuals (the first M in the method's name). The estimated scale is then held constant whilst a close-by M-estimate of the parameters is located (the second M) \cite{yohai1987}. This provides robustness both to outliers and to data points with high leverage. Further robustness is provided by using Tukey's bisquare objective function \cite{mosteller1977}: instead of minimizing the sum of squared residuals, we minimize the sum of a function of the residuals that is capped at a maximum value for each residual. This means that an outlier in the regression analysis has the same contribution to the objective function no matter how extreme the outlier is.

If the objective function of the standardized residuals $r_j$ in the regression is $\sum_j \rho(r_j)$, then ordinary least squares regression minimizes the sum of the square of the residuals, $\rho(r_j) = r_j^2$. In Tukey's bisquare objective function,
\begin{equation}
\rho(r_j) = \begin{cases}\frac{c^2}{6} \left\{1-\left[1-\left(\frac{r_j}{c}\right)^2\right]^3\right\} \quad &\mbox{if } |r_j|<c \\
                         \frac{c^2}{6}                                                                \quad &\mbox{if } |r_j| \ge c. \end{cases}
\end{equation}
The value of the tuning parameter $c$ is chosen as 1.548 to provide a high breakdown point in the S-estimation step, and as 4.685 to provide an efficient estimator in the M-estimation steps. This method for robust regression is the default choice implemented by the \texttt{lmrob} command in the R package \emph{robustbase} \cite{robustbase}. Robust regression can replace standard regression in both the IVW and MR-Egger methods.

\subsection{Penalization of weights}
Another way of providing additional robustness is to penalize the weights of candidate instruments with heterogeneous ratio estimates in the weighted regression model. This could be achieved in many ways; we propose an approach using Cochran's Q statistic as a measure of heterogeneity. For the IVW method:
\begin{equation}
Q = \sum_j Q_j = \sum_j \sigma^{-2}_{Yj} (\hat{\beta}_{Yj} - \hat{\theta} \hat{\beta}_{Xj})^2
\end{equation}
where $\hat{\theta}$ is here taken as the IVW estimate. The Q statistic has an approximate $\chi^2_{J-1}$ distribution under the null hypothesis that all candidate instruments satisfy the instrumental variable assumptions; the components of the Q statistic for each candidate instrument ($Q_j$) approximately have $\chi^2_{1}$ distributions. So as not to distort the majority of weights, we propose penalization using the one-sided upper p-value (denoted $q_j$) on a $\chi^2_{1}$ distribution corresponding to $Q_j$, by multiplying the weight ($\se(\hat{\beta}_{Yj})^{-2}$) by $\min(1, 20 q_j)$. The same downweighting factor has previously been used for weights in the median-based method to give a penalized weighted median estimate \cite{bowden2015median}. For the median-based methods, we replace the IVW estimate by the relevant median estimate in the calculation of the Q statistic.

For the MR-Egger method, we consider a Q statistic equivalent to the residual sum of squares from the weighted regression, which has an approximate $\chi^2_{J-2}$ distribution if the MR-Egger regression model is correct \cite{rucker2010}:
\begin{equation}
Q = \sum_j Q_j = \sum_j \sigma^{-2}_{Yj} (\hat{\beta}_{Yj} - \hat{\theta}_{0} - \hat{\theta}_{1} \hat{\beta}_{Xj})^2.
\end{equation}

If most candidate instruments are valid instrumental variables, then robust regression in either the IVW or MR-Egger method should give a consistent causal estimate asymptotically as the sample size increases, as the association estimates from valid instruments should all align on a straight-line through the origin \cite{kang2014} (the value of ``most'' depends on the breakdown point of the robust regression method). Equally, provided that the causal estimate in the Q statistic is close to the true causal effect, penalization of weights should downweight the contribution of invalid instruments to zero asymptotically as the sample size increases. If the pleiotropic effects of candidate instruments are independently distributed from their associations with the exposure (a population version of the InSIDE assumption), then the weighted correlation between the pleiotropic effects and associations with the exposure should tend to zero asymptotically as the number of candidate instruments increases for all choices of weights, and should be zero on average for a random choice of candidate instruments. This means that penalization of weights should not affect the consistency of the MR-Egger method under the population InSIDE assumption, nor the use of any robust regression method that is equivalent to varying the weights. This provides some motivation that these extensions should provide reasonable estimates in large samples. However, it is unclear what will happen if some variants satisfy the InSIDE assumption, but others do not.

\subsection{L1 penalization}
An alternative estimation method is to fit a separate intercept coefficient for each candidate instrument, and to use penalization to identify the model. In the standard MR-Egger method, a single intercept term is estimated, representing the average pleiotropic effect. Weighted linear regression in the MR-Egger method minimizes the following expression:
\begin{equation}
\sum_j \se(\hat{\beta}_{Yj})^{-2} (\hat{\beta}_{Yj} - \theta_0 - \theta_1 \hat{\beta}_{Xj})^2
\end{equation}
We propose replacing the $\theta_0$ with $\theta_{0j}$, and adding an L1-penalty term:
\begin{equation}
\sum_j \se(\hat{\beta}_{Yj})^{-2} (\hat{\beta}_{Yj} - \theta_{0j} - \theta_1 \hat{\beta}_{Xj})^2 + \lambda \sum_j |\theta_{0j}|
\end{equation}
where $\lambda$ is a tuning parameter. If $\lambda = 0$, then all the candidate instruments are allowed to be pleiotropic, and the model is not identified. If $\lambda = \infty$, then this is equivalent to the IVW method using all candidate instruments, as the pleiotropic effects are forced to take the value zero -- in effect, all candidate instruments are assumed to be valid instruments. As the value of $\lambda$ decreases, the number of candidate instruments for which $\alpha_j \neq 0$ increases, and these candidate instruments are allowed to be pleiotropic. An advantage of L1 penalization over other penalization options is the sparsity property -- some coefficients are shrunk to zero, representing invalid instruments. Additionally, it can be shown that the estimate of $\theta_1$ in L2 penalization does not depend on the value of the tuning parameter (see Web Appendix A.2).

Once the value of $\lambda$ is determined, we perform the IVW method using all candidate instruments that are determined to be valid instruments (that is, for all $j$ such that $\hat{\theta}_{0j} = 0$). This provides a causal estimate and a standard error.

\section{Simulation study}
\label{sec:simul}
We perform a simulation study to compare the bias and coverage properties of estimates from different methods:
\begin{itemize}
\item standard linear regression without and with an intercept term using inverse-variance weights as in equation~(\ref{eq:egger}) -- this is equivalent to the IVW and MR-Egger methods respectively;
\item robust linear regression (MM-estimation with bisquare objective function) without and with an intercept term using inverse-variance weights;
\item standard linear regression without and with an intercept term using penalized inverse-variance weights;
\item robust linear regression without and with an intercept term using penalized inverse-variance weights;
\item L1 penalization using various approaches for selecting the tuning parameter;
\item simple, weighted, and penalized weighted median estimates (for comparison).
\end{itemize}
We investigate whether these methods give reasonable inferences (in particular, maintain nominal Type 1 error rates under the causal null hypothesis [$\theta = 0$], but have reasonable power under the alternative) in realistic scenarios. Robust regression is implemented using the \texttt{lmrob} command from the \emph{robustbase} package in R \cite{r312} with the \texttt{method = "MM"} option \cite{robustbase}.

\subsection{Data-generating model}
The data-generating model for the simulation study is as follows:
\begin{align}
U_{i}  &= \sum^{J}_{j=1}   \phi_{j} G_{ij}                        + \epsilon_{Ui} \label{eq:datgen} \\
X_{i}  &= \sum^{J}_{j=1} \gamma_{j} G_{ij}                + U_{i} + \epsilon_{Xi} \notag \\
Y_{i}  &= \sum^{J}_{j=1} \alpha_{j} G_{ij} + \theta X_{i} + U_{i} + \epsilon_{Yi} \notag \\
G_{ij} &\sim \mbox{Binomial}(2, 0.3) \mbox{ independently for all $j = 1, \ldots, J$} \notag \\
\epsilon_{Ui}, \epsilon_{Xi}, \epsilon_{Yi} &\sim \mathcal{N}(0, 1) \mbox{ independently} \notag
\end{align} %
for participants indexed by $i = 1, \ldots, N$, and candidate instruments indexed by $j = 1, \ldots, J$. The candidate instruments $G_{j}$ are simulated to be equivalent to genetic variants that are single nucleotide polymorphisms in Hardy--Weinberg equilibrium with minor allele frequency 0.3. The variable $U$ is a confounder in the relationship between the exposure and the outcome, and is assumed to be unmeasured. The error terms $\epsilon_{Ui}$, $\epsilon_{Xi}$, and $\epsilon_{Yi}$ were each drawn independently from standard normal distributions. The causal effect of the exposure on the outcome was either $\theta = 0$ (null causal effect) or $\theta = 0.1$ (positive causal effect). The effects of the candidate instruments on the exposure ($\gamma_j$) were drawn from a uniform distribution between 0.03 and 0.1. The direct effects of a candidate instrument (genetic variant) on the outcome ($\alpha_j$) and the effects of the candidate instruments on the confounder ($\phi_j$) were set to zero if the candidate instrument was a valid instrumental variable; for candidate instruments that were invalid instrumental variables:
\begin{itemize}
\item In Scenario 2 (direct effects average to zero -- balanced pleiotropy, population InSIDE satisfied), the $\alpha_j$ parameters were drawn from a uniform distribution between $-0.1$ and 0.1, and the $\phi_j$ were taken as 0.
\item In Scenario 3 (direct effects do not average to zero -- directional pleiotropy, population InSIDE satisfied), the $\alpha_j$ parameters were drawn from a uniform distribution between 0 and 0.1, and the $\phi_j$ were taken as 0.
\item In Scenario 4 (direct effects operate via confounder and hence do not average to zero -- directional pleiotropy, InSIDE not satisfied), the $\phi_j$ parameters were drawn from a uniform distribution between $-0.1$ and 0.1, and the $\alpha_j$ were taken as 0.
\end{itemize}
In Scenario 1, all candidate instruments are taken to be valid instruments. In Scenarios 2 to 4, each candidate instrument was determined to be a valid or an invalid instrumental variable based on a Bernoulli trial with the probability of being invalid set to 0.1, 0.2, and 0.3. Although we only consider scenarios with (on average) up to 30\% invalid instruments, the pleiotropic effects of invalid instruments are fairly large. The maximal indirect association of a candidate instrument with the outcome via the exposure with a positive causal effect is $0.1 \times 0.1 = 0.01$ (if $\gamma_j = 0.1$), whereas the maximal direct (pleiotropic) effect is 0.1 (if either $\alpha_j = 0.1$ or $\phi_j = 0.1$),

A total of 10\thinspace000 simulated datasets were generated for $N = 20\thinspace000$ participants and $J = 25$ candidate instruments. A `two-sample' setting was assumed in which associations of the candidate instruments with the exposure were estimated in $N$ participants, and associations with the outcome in a separate sample of $N$ participants. Results obtained in a one-sample setting in which the associations with the exposure and with the outcome are obtained in the same individuals are given in Web Appendix A.3. Only the summarized data, that is the estimated univariable associations of the candidate instruments with the exposure and with the outcome, and their standard errors, were used by the analysis methods. The average proportion of variance in the exposure explained by the candidate instruments ($R^2$ statistic) was 2.5\% (2.8\% in Scenario 4), and the average F statistic was 20.5 (23.3 in Scenario 4). 

Six strategies were undertaken for choosing the value of the tuning parameter $\lambda$ in the L1 penalization method. We set $\lambda = 1$, $\lambda = 2$, and $\lambda = 3$, to compare the performance of the method with different choices of the tuning parameter. Fourthly, we used leave-one-out cross-validation of the likelihood function. Fifthly, we used a grid search to pick the value of $\lambda$ that gave the causal estimate closest to zero; estimates were calculated for $\lambda = 0.1, 0.2, \ldots, 4.9, 5.0, 5.2, 5.4, \ldots, 9.8, 10.0$. Finally, we used a heterogeneity criterion to determine how many variants to include in the model; we increased the value of the tuning parameter by the same increments as in the grid search, stopping when the residual standard error in the regression model was above 1 and the next value of $\lambda$ increased the residual standard error by an increment of more than $\chi^2_1 (0.95)/(J_{inc}-1)$, where $\chi^2_1 (0.95)$ is the upper 95th percentile of a chi-squared distribution on 1 degree of freedom, and $J_{inc}$ is the number of candidate instruments included in the model. This is motivated by Cochran's Q heterogeneity statistic (equal to the residual standard error multiplied by the number of candidate instruments less 1) having a $\chi^2_{J-1}$ distribution under the null hypothesis that all candidate instruments are estimating the same causal parameter.

\subsection{Results}
\begin{figure}[htbp]
\begin{center}
\includegraphics[width=14cm]{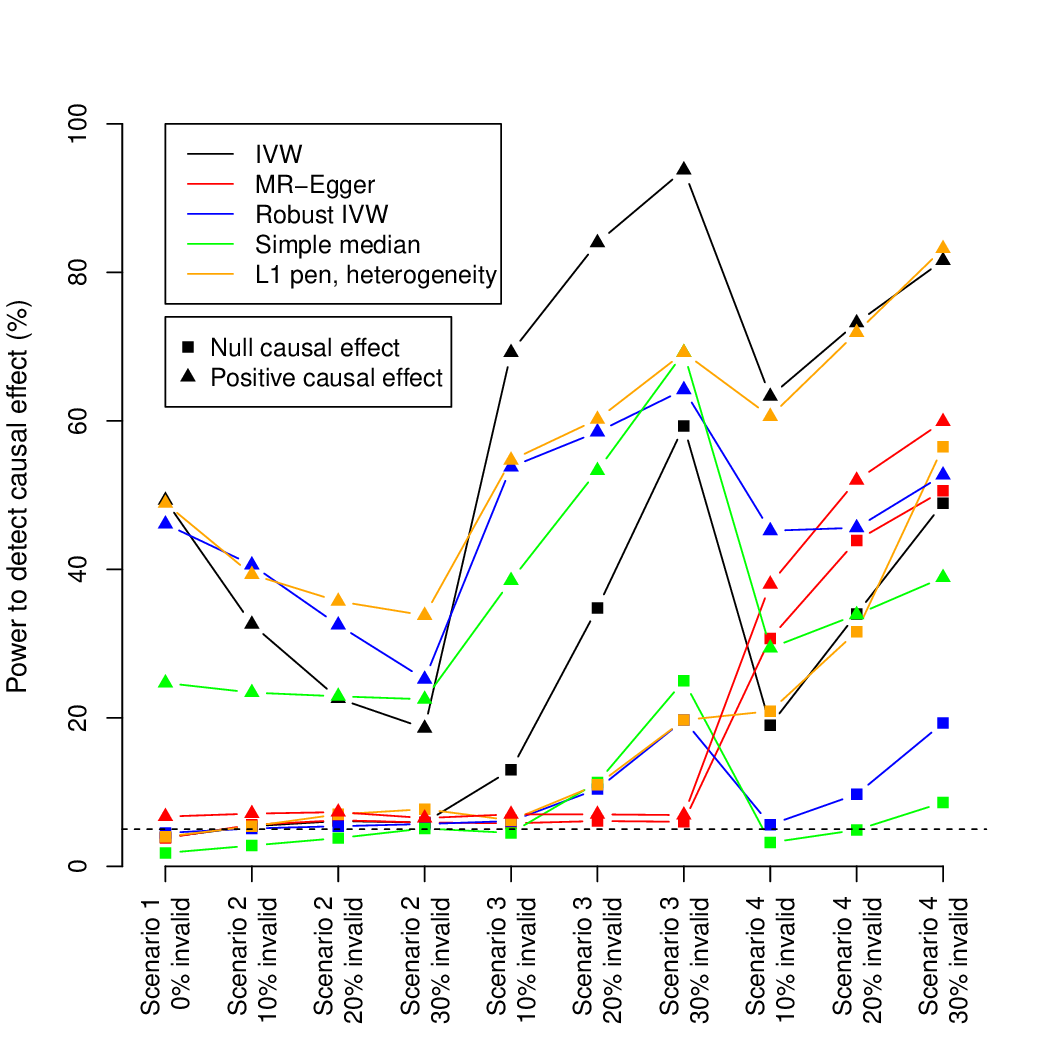}
\end{center}
\caption{Power to detect a causal effect (equivalent to Type 1 error rate with null causal effect) for selected methods in each scenario. The dashed line is at 5\%; the nominal power expected with a null causal effect.} \label{robustcover}
\end{figure}

Results from the simulation study are provided in Table~\ref{simresults.1a} (Scenario 1 only, all methods), Table~\ref{simresults.2a} (Scenarios 2-4, weights not penalized), Table~\ref{simresults.2b} (Scenarios 2-4, penalized weights), and Table~\ref{simresults.2c} (Scenarios 2-4, L1 penalization methods). Table~\ref{simresults.1a} displays the mean estimate across simulations, standard deviation of estimates, mean standard error of estimates, and the empirical power to detect a causal effect (the proportion of simulations where the 95\% confidence interval [estimate $\pm$ 1.96 standard errors] excluded the null). With a null causal effect, power to detect a causal effect is the same as the Type 1 error rate, and the expected power is 5\%. In Tables~\ref{simresults.2a}, \ref{simresults.2b} and \ref{simresults.2c}, the mean standard error of estimates is omitted (the pattern of results for the mean standard error was similar to that in Scenario 1 except as noted below). In some of the simulations, the robust regression method did not report a standard error (less than 1\% in all cases, except up to 2.5\% for the robust method with an intercept in Scenario 4); the number of simulations that failed to report a standard error is given in Table~\ref{simresults.1a} for Scenario 1, and in Web Table~\ref{simresults.na} for other scenarios. Simulations were not excluded from the results if a standard error was not reported (except for the calculation of the mean standard error); power calculations include these simulations as if the standard error estimate is infinite. The Monte Carlo standard error (the uncertainty due to the limited number of simulations considered) for the power was 0.2\% with a null effect, and between 0.2\% and 0.5\% with a positive causal effect. A graph illustrating the coverage rates for a limited selection of methods is provided as Figure~\ref{robustcover}.

\textbf{Scenario 1 (Table~\ref{simresults.1a}):} When all candidate instruments were valid instruments, all methods provided unbiased estimates under the null, with Type 1 error rates close to or below the nominal significance level of 5\%. The standard deviation of estimates was slightly below the mean standard error of estimates for all methods, with differences most marked for the median-based methods. This difference suggests that methods may be slightly conservative in their inferences. In terms of precision of the causal estimate, regression methods without an intercept (including the IVW method) and L1 penalization methods were the most precise, followed closely by the median-based methods, while regression methods with an intercept (including the MR-Egger method) were the least precise. Differences in precision between the standard, robust and penalized methods were slight. The exception was the L1 penalization method taking the minimal estimate, which gave conservative inference and less variable estimates, particularly under the null.

With a positive causal effect, differing precisions of the causal estimate were also evidenced by the marked differences in power to detect a causal effect. The power for the regression methods including an intercept term was barely above 5\%. While precision of the causal estimate for the IVW method depends on the proportion of variance in the exposure explained by the candidate instruments, precision of the causal estimate for MR-Egger depends on the variability between the instrument--exposure associations \cite{bowden2016}. If all candidate instruments have exactly the same magnitude of association with the exposure, then the MR-Egger estimate is undefined. The MR-Egger estimate will always be less precise than the IVW estimate, but the difference in precision will depend on whether the instrument--exposure associations for different candidate instruments are similar to each other or not.

While there was some attenuation towards the null with a positive causal effect for all the methods (except for the simple median method) due to uncertainty in the associations of the candidate instruments with the exposure, this was minimal for the IVW and other methods with no intercept, but substantial for the MR-Egger and other methods with an intercept. This attenuation is a known phenomenon called finite-sample bias (also known as weak instrument bias \cite{burgess2010weak}). Bias in the two-sample setting acts towards the null \cite{pierce2013} and is related to regression dilution bias \cite{frost2000}; it arises due to measurement error in the independent variable in a regression model. Relative bias of the IVW estimate is around 1/$F$, where $F$ is the expected value of the F statistic from regression of the exposure on the IVs (here, 1/$F$ $\approx 1/20 = 5\%$, similar to the observed attenuation in the mean IVW estimates) \cite{staiger1997}; whereas attenuation of the MR-Egger estimate is approximately equal to the $I^2$ statistic from meta-analysis of the weighted associations with the exposure $\hat{\beta}_{Xj} \se(\hat{\beta}_{Yj})^{-1}$ with standard errors $\se(\hat{\beta}_{Xj}) \se(\hat{\beta}_{Yj})^{-1}$ \cite{bowden2016}. The $I^2$ statistic is large when the candidate instruments have a wide spread of associations with the exposure or their associations are precisely estimated, and small when their associations with the exposure are imprecisely measured or all similar. In the simulation, the average value of the $I^2$ statistic was 60.1\%. This bias can be corrected using the Simulation Extrapolation (SIMEX) method \cite{cook1994, bowden2016}, although this was not computationally feasible in the simulation setting. While measurement error in the exposure can lead to inflation of the intercept term in the MR-Egger method, in this case the 95\% confidence interval for the intercept term excluded zero for MR-Egger in 4.7\% of simulations -- close to the expected nominal 5\% level, indicating that over-rejection of the null hypothesis for the MR-Egger intercept test was not evident in this example.

\begin{center}
[Table~\ref{simresults.1a} should appear about here.]
\end{center}

\textbf{Scenario 2, 3 and 4, non-penalized weights (Table~\ref{simresults.2a}):} Mean estimates in Scenario 2 (balanced pleiotropy, InSIDE satisfied) were unbiased with a null causal effect for all methods. With a positive causal effect, mean estimates were similar to those in Scenario 1: close to unbiased for most methods, but with severe attenuation for regression methods with an intercept term. However, there were marked differences in the precision of estimates compared with Scenario 1. Out of previously proposed methods, estimates from the median-based methods were more precise than those from the IVW method, although this did not translate into greater power with a positive causal effect when only 10\% of candidate instruments were not valid instruments. However, the greatest power was obtained by the robust regression method with no intercept. Although the power of the MR-Egger method and other regression methods with an intercept was low, the use of robust regression did reduce the standard deviation and mean standard error of estimates.

Scenario 3 (directional pleiotropy, InSIDE satisfied) demonstrates the value of the MR-Egger and related methods estimating an intercept for providing robust inferences under the InSIDE assumption. While estimates from other methods (particularly the IVW method) were biased under the null, mean estimates from regression methods with an intercept term were close to unbiased, and Type 1 error rates were close to nominal levels. But again, these methods were unable to identify the presence of a causal effect with reasonable power, and mean estimates were substantially attenuated. More seriously, in Scenario 4 (directional pleiotropy, InSIDE not satisfied), the MR-Egger method performed much worse than the IVW method, with mean estimates far more biased and larger Type 1 error rates. While there was some improvement using robust regression with an intercept term when there were few invalid instruments, there was still substantial bias and Type 1 error inflation, as well as even less precise estimates compared with the MR-Egger method when there were many invalid instruments. The MR-Egger and related methods are highly sensitive to the validity of the InSIDE assumption. As the InSIDE assumption is not testable, this is a major limitation of these methods.

In contrast, while the median-based methods and robust regression method without an intercept had bias in mean estimates and inflated Type 1 error rates, rates were substantially below those for the IVW method. In particular, Type 1 error rates were close to 10\% or below for the simple median and robust regression without an intercept methods in Scenarios 3 and 4 with up to 20\% invalid instruments, and for the simple median method in Scenario 4 with 30\% invalid instruments. The weighted median method was particularly poor in Scenario 4; the data-generating mechanism meant that the invalid instruments received more weight in the analysis than the valid instruments as they had greater associations with the exposure (via an additional association with the confounder). The median-based methods and robust regression without an intercept also had reasonable power to detect a causal effect when present. 

\begin{center}
[Table~\ref{simresults.2a} should appear about here.]
\end{center}

\textbf{Scenarios 2, 3 and 4, penalized weights (Table~\ref{simresults.2b}):} The use of penalized weights generally led to more precise causal estimates, and Type 1 error rates were somewhat improved in Scenarios 3 and 4 for the IVW and weighted median methods. However, Type 1 error rates for the penalized methods generally exceeded nominal levels in all scenarios, especially when 20\% or more candidate instruments were invalid. A particular cause for concern is the inflated Type 1 error rates in Scenario 2, which did not occur with unpenalized weights. The reason seems to be that the heterogeneity between the estimates from candidate instruments was underestimated, and hence there was underestimation of the uncertainty in the causal estimate. This highlights a danger that penalization of weights can lead to overconfidence in making inferences, as evidence that points in a different direction is downweighted in the analysis. Penalization of weights did not seem to be a worthwhile strategy for controlling Type 1 error rates in this simulation study.

\begin{center}
[Table~\ref{simresults.2b} should appear about here.]
\end{center}

\textbf{Scenarios 2, 3 and 4, L1 penalization method (Table~\ref{simresults.2c}):} Similarly, although Type 1 error rates for the implementations of the L1 penalization method were improved compared with the IVW method in Scenarios 3 and 4, there was slight overprecision in Scenario 2, and Type 1 error rates were consistently greater than nominal levels. Estimates using the tuning parameter ($\lambda$) chosen by cross-validation had the largest Type 1 error rates and the most variable estimates, but did not always have the greatest power to detect a causal effect. This suggests that cross-validation tended to include too few candidate instruments in the causal analysis, and in Scenarios 3 and 4, often chose the wrong candidate instruments. Estimates using the heterogeneity criterion to choose the values of $\lambda$ produced better inferences than by simply choosing a fixed value of $\lambda$ in Scenarios 2 and 3, but worse in Scenario 4. When the value of $\lambda$ was chosen to give the minimal causal estimate, Type 1 error rates were conservative in Scenario 2, and in Scenario 3 with up to 20\% invalid instruments. However, power to detect a causal estimate was also considerably lower.

\textbf{Supplementary analyses (not done yet for L1 penalization):} This simulation was repeated in a one-sample setting in which associations of the candidate instruments with the exposure and with the outcome were obtained in the same sample of 20\thinspace000 individuals for the methods using non-penalized weights. Results are displayed in Web Table~\ref{simresults.1one} (Scenario 1) and Web Table~\ref{simresults.2one} (Scenarios 2 to 4). Bias in the direction of the observational association was observed in all methods except for the simple median method (which remained unbiased in Scenarios 1 and 2). However, the bias of the MR-Egger method was greater and more severe than that of the IVW method: in Scenario 1 with a null causal effect, the mean estimates were 0.024 for the IVW method and 0.173 for the MR-Egger method, and Type 1 error rates were 6.8\% and 27.2\% respectively. The rejection rate of the MR-Egger intercept test was also inflated (23.5\% with a null causal effect, 20.3\% with a positive causal effect). The one-sample setting is another case where the MR-Egger method performs poorly.

The simulation was also repeated in a two-sample setting with only 10 candidate instruments, to observe whether the robust methods were able to operate well with fewer instruments to detect violations of the instrumental variables assumptions. Results for Scenarios 2 to 4 are presented in Web Table~\ref{simresults.2ten}. Power to detect a causal effect was generally much lower, but otherwise similar results were observed.

Finally, Table~\ref{simresults.3} shows the proportion of datasets for the original simulation study rejecting the causal null using both the simple median and robust regression method with no intercept (robust IVW), and the empirical power of the MR-Egger intercept test for detecting directional pleiotropy and/or violations of the InSIDE assumptions. The combination of the simple median and robust IVW methods generally provided conservative inferences, with Type 1 error rates close to or below nominal levels except in Scenario 3 with 30\% invalid instruments. This suggests that multiple robust methods could be used as sensitivity analyses in practice to better control Type 1 error rates. The MR-Egger intercept test is a test of directional pleiotropy and/or violation of the InSIDE assumption: as expected, rejection rates were around 5\% in Scenarios 1 and 2, and greater in Scenarios 3 and 4. This suggests that, even if the MR-Egger estimate is unreliable, the method may be useful for detecting in which cases the IVW method is likely to be biased.

\begin{center}
[Table~\ref{simresults.3} should appear about here.]
\end{center}

\section{Applied example: causal effect of body mass index on schizophrenia risk}
\label{sec:appliedex}
\begin{figure}[htbp]
\begin{center}
\includegraphics[width=10cm]{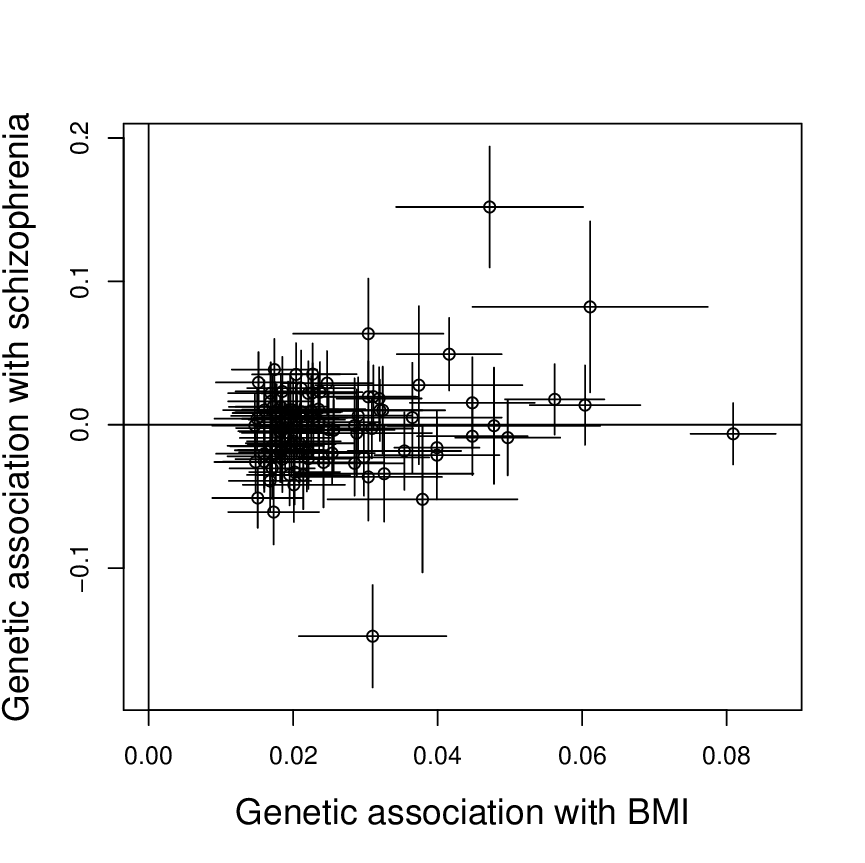}
\end{center}
\caption{Estimated genetic associations and 95\% confidence intervals with body mass index (BMI, standard deviation units) and with schizophrenia risk (log odds ratios) for 97 genetic variants.} \label{assocs}
\end{figure}

As an applied example to illustrate the methods, we considered the causal effect of body mass index (BMI) on schizophrenia risk. Individuals with schizophrenia generally have higher incidence of obesity than the general population \cite{coodin2001}, although the relationship is thought to arise from the effect of anti-psychotic medicine on BMI (reverse causation) rather than as a causal effect of BMI \cite{allison1999}. We use 97 genetic variants previously demonstrated to be associated with BMI at a genome-wide level of significance by the Genetic Investigation of Anthropometric Traits (GIANT) consortium \cite{locke2015}. Associations with the exposure were taken from univariable linear regression analyses in up to 339\thinspace224 European-descent individuals from the GIANT consortium \cite{locke2015}; associations with the outcome were taken from univariable logistic regression analyses in around 9000 European-descent cases and 8000 controls from the Psychiatric Genomics Consortium \cite{smoller2013}. The 97 genetic variants explain about 2.7\% of the variance in BMI. Both sets of genetic associations have previously been made publicly available, and the association estimates can be obtained using the PhenoScanner tool at \url{http://phenoscanner.medschl.cam.ac.uk/}; they are also displayed visually in Figure~\ref{assocs}. The graph indicates that there are several genetic variants that are clear outliers in their associations with schizophrenia, suggesting potential pleiotropy. The $I^2$ statistic for the weighted genetic associations with the exposure was 88.8\%, suggesting that attenuation of the MR-Egger and other methods that estimate an intercept should not be severe.

\begin{center}
[Figure~\ref{assocs} should appear about here.]
\end{center}

Estimates and 95\% confidence intervals are provided in Table~\ref{exresults}. Random-effects models were used in all analyses. Each estimate represents the log odds ratio for schizophrenia per 1 standard deviation increase in BMI. Although all estimates are compatible with the null, there is a wide variation in the standard errors of estimates. Using non-penalized weights, a similar pattern of results was seen as in the simulation analyses of Scenario 2: the robust method with no intercept giving the most precise estimate, followed by the median-based methods, with the MR-Egger method far behind. The use of penalized weights led to large improvements in precision for all except the median-based methods, indicating that although penalization of weights did not seem to add robustness to results in the simulation study, it may have a role in improving the precision of results in cases like this where there are genetic variants that are clear outliers. In the IVW method, the use of penalized weights reduced the residual standard error from 2.14 to 1.12, only slightly above the value of 1 that would be expected in the absence of heterogeneity. In an applied setting, the genetic variants that are downweighted in the analysis should be examined for pleiotropy to determine whether their omission from the analysis is reasonable.

\begin{figure}[htbp]
\begin{center}
\includegraphics[width=14cm]{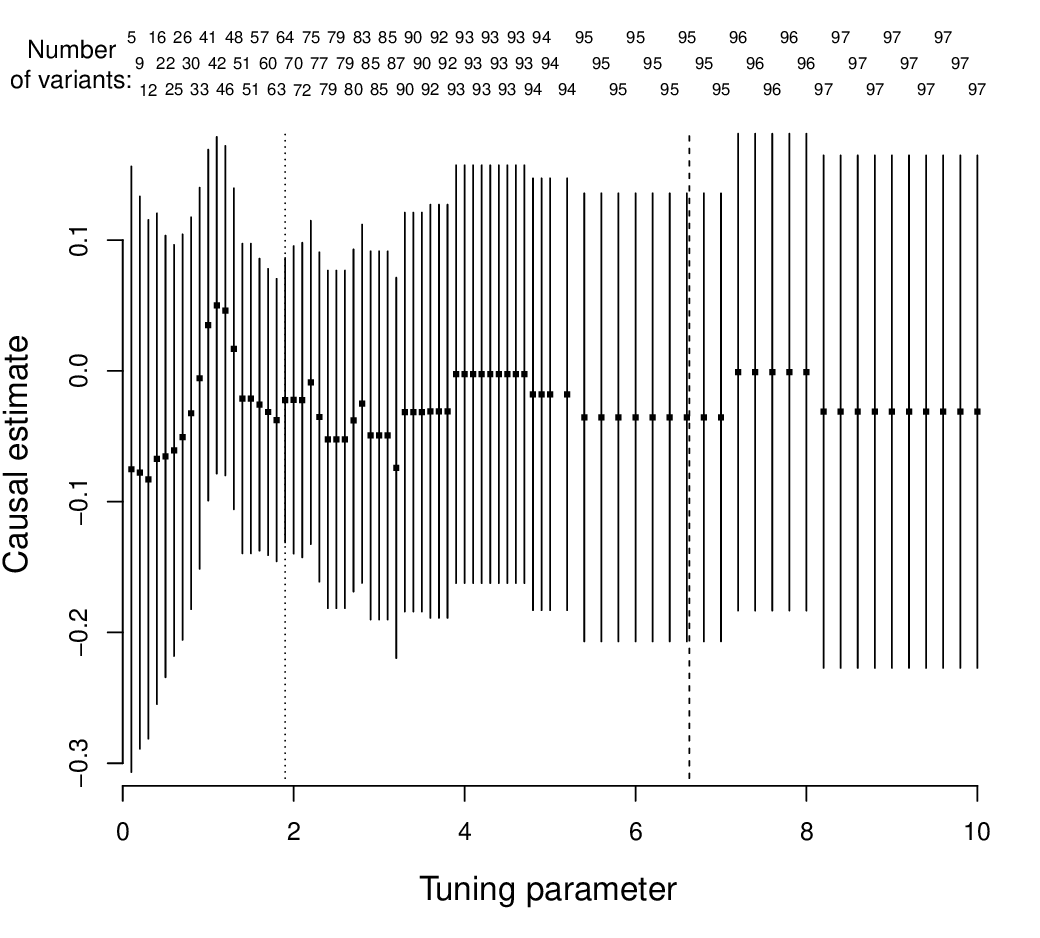}
\end{center}
\caption{Causal effect estimates (point estimate and 95\% confidence interval) for a range of values of the tuning parameter in applied example (BMI on schizophrenia). The number of genetic variants included in each analysis is also displayed. The dotted line at $\lambda = 1.9$ is the value of the tuning parameter chosen by the heterogeneity criterion. The dashed line at $\lambda = 6.63$ is the value chosen by cross-validation.} \label{l1tune}
\end{figure}

L1 penalization estimates for a range of values of the tuning parameter are displayed in Figure~\ref{l1tune}. Using the heterogeneity criterion, the value of the tuning parameter was $\lambda = 1.9$ and 64 genetic variants were included in the analysis. Using cross-validation, the value of the tuning parameter was much larger at $\lambda = 6.63$ and 95 of the 97 genetic variants were included in the analysis. The heterogeneity criterion almost chose the value of $\lambda$ corresponding to the most precise causal estimate ($\lambda = 1.8$ gave a slightly more precise estimate), and a more precise estimate than from any other method. However, as can be seen in Figure~\ref{l1tune}, causal estimates were fairly similar in magnitude whatever value of $\lambda$ was chosen, even for $\lambda = 0.1$ when only 5 genetic variants were included in the analysis. While in the simulation study, a strategy was required to choose the value of $\lambda$, in practice causal estimates can be compared using a range of values of the tuning parameter.

This applied example illustrates that in addition to providing additional confidence in the robustness of findings from a conventional analysis, the methods introduced in this paper have the potential to improve the efficiency of Mendelian randomization estimates.

\begin{center}
[Table~\ref{exresults} should appear about here.]
\end{center}

\section{Discussion}
\label{sec:discuss}
In this paper, we have introduced three extensions to instrumental variable methods to downweight the influence of candidate instruments with heterogeneous causal estimates, with the aim of providing more robust estimates in Mendelian randomization investigations. A summary of the methods presented in this paper is provided in Table~\ref{summary}.

\begin{center}
[Table~\ref{summary} should appear about here.]
\end{center}

While the robust and the penalized versions of MR-Egger have desirable theoretical properties, in our simulation study neither method was able to reliably detect the presence of a causal effect of moderate size with reasonable power. Additionally, both these and the original MR-Egger method were highly sensitive to violations of the InSIDE assumption. The MR-Egger intercept test was able to detect scenarios in which the IVW method gave biased estimates, although power was moderate at best. The two methods that had the best performance across the range of scenarios considered in terms of Type 1 error rate and power were the robust version of the IVW method and the simple median method. Although Type 1 error rates were inflated over nominal levels for all methods in at least one scenario, improvement over the standard inverse-variance weighted method was considerable.

If alternative parameters or scenarios were chosen in the simulation study, then different results might have been observed. For example, if candidate instruments had substantially different strengths (and validity of the candidate instruments did not depend on instrument strength, as in Scenario 4), then the weighted median method may have been preferable to the simple median method, and the loss of power in the MR-Egger method compared with the IVW method would have been less severe. Alternatively, if simulations were conducted in a scenario where 100\% of the candidate instruments were invalid but they satisfied the InSIDE assumption, then the MR-Egger method would have fared better; likewise if the magnitude of the causal effect was greater (hence the MR-Egger method would have had improved power to detect a causal effect), or if the sample size for the genetic associations with the exposure increased (hence the MR-Egger estimates would have been less attenuated). Hence, the conclusion from this work should not be to promote one method for Mendelian randomization analysis to the exclusion of others, but rather to emphasize the need for multiple sensitivity analyses that make different sets of assumptions. The robust version of the IVW method seems to be a worthwhile sensitivity analysis method in addition to other robust methods previously proposed (such as simple and weighted median, and MR-Egger \cite{burgess2016robust}). The use of penalized weights may be worthwhile to improve precision if a small number of candidate instruments have clearly heterogeneous causal estimates (as demonstrated in the applied example), but the approach is unlikely to lead to robust inferences if several candidate instruments are not valid. Similarly, L1 penalization can improve precision when causal estimtes are heterogeneous, and the approach gave improved inferences over the IVW method when the InSIDE assumption was satisfied. While it is not clear how to best choose the tuning parameter in an automated way for a simulation analysis, in an applied example estimates can be reported for a range of values of this parameter. Additionally, if we had considered weaker pleiotropic effects in the simulation study, Type 1 error inflation would have been less pronounced.

\subsection{Linearity and homogeneity assumptions}
In the specification of the analysis models, we have assumed linearity and homogeneity (no effect modification) of the causal effect of the exposure on the outcome, and of the associations of the candidate instruments with the exposure and with the outcome. These assumptions are not necessary to identify a causal effect; weaker assumptions can be made \cite{swanson2013} (such as monotonicity of the causal effect \cite{imbens1994} or a weaker version of the homogeneity assumption for the causal effect \cite{robins1989g, hernan2006}). If the linearity and homogeneity assumptions are violated, then the causal estimate using a single instrumental variable is a valid test of the null hypothesis that the exposure does not have a causal effect on the outcome \cite{didelez2007}; this also applies to the causal estimate from the IVW method using multiple instruments, as this is a linear combination of the causal estimates from the individual instruments \cite{burgess2015scoretj}. Hence, even when the linearity and homogeneity assumptions are violated, the methods proposed in this paper can still be used for the assessment of the causal null hypothesis (does the exposure have a causal effect on the outcome?), even if the estimate does not have a literal interpretation \cite{burgess2015beyond}.

Additionally, while the linearity and homogeneity assumptions are stringent, genetic variants tend to have small effects on the exposure and outcome. This means that linearity and homogeneity may not be unreasonable assumptions in an applied Mendelian randomization investigation. Linearity and homogeneity in the genetic associations are not required across the whole distribution of the exposure and the outcome, but simply in the range of values predicted by the genetic variants.

\subsection{Alternative robust methods}
Several other methods have been developed for robust estimation using instrumental variables. Koles\'{a}r et al.\ proposed a method within the framework of k-class estimators with a 100\% breakdown level under the InSIDE assumption \cite{kolesar2014}. Kang et al.\ proposed a method using individual-level data based on penalized regression for detecting and accounting for invalid instruments that provides a consistent estimate of causal effect if at least 50\% of the candidate instruments are valid using L1 penalization to downweight the contribution to the analysis of candidate instruments that have heterogeneous causal estimates \cite{kang2014}. Han proposed a similar penalized estimator within the generalized method of moments framework, again with a 50\% breakdown level \cite{han2008}. These ideas were developed further by Windmeijer et al.\ \cite{windmeijer2016}; several of the choices made here relating to the L1 penalization method (such as the decision to use the method to identify invalid instruments and to obtain the causal estimate using only the valid instruments, referred to by Windmeijer et al.\ as a `post-lasso' estimator, and the decision to explore a heterogeneity criterion for selecting the tuning parameter) were guided by that paper. However, each of these methods requires individual-level data, limiting their applicability to applied Mendelian randomization investigations.

\subsection{Conclusion}
We have shown that it is difficult to find methods that give robust causal inferences with invalid instruments. Even in the examples with moderate numbers of invalid instruments considered in this paper, all methods had inflated Type 1 error rates in at least one scenario. Nevertheless, although the methods we have proposed are far from perfect, they have much improved Type 1 error rates compared with the conventional IVW method and the recently introduced MR-Egger method in scenarios where the InSIDE assumption fails to hold.

We have demonstrated that using multiple methods for instrumental variable analysis (particularly methods that provide consistent estimates under different assumptions) can provide more reliable inferences for Mendelian randomization investigations. A causal conclusion is more plausible in cases where multiple methods suggest a causal effect. We suggest that the IVW method using robust regression is a worthwhile method to apply in addition to previously proposed methods (in particular the simple median method), and that the use of penalized weights and L1 penalization may be valuable in some situations.

\subsection*{Acknowledgements}
Stephen Burgess is supported by the Wellcome Trust (grant number 100114). Jack Bowden is supported by a Methodology Research Fellowship from the Medical Research Council (grant number MR/N501906/1). Frank Dudbridge is supported by the Medical Research Council (grant number K006215). Simon G.\ Thompson is supported by the British Heart Foundation (grant number CH/12/2/29428).

\bibliographystyle{wileyj} 
\bibliography{masterref}

\clearpage
\subsection*{Tables}

\begin{table}[htbp] 
\begin{minipage}{\textwidth}
\begin{center}
\begin{small}
\centering
\begin{tabular}[c]{c|ccccc}
\hline
                                  & \multicolumn{5}{c}{Scenario 1}                         \\
Method                            &  Mean    & SD     &  Mean SE  &  Power       &  NA
   \footnote{Number of the 10\thinspace000 simulations that failed to report a standard error.} \\
\hline
\multicolumn{6}{c}{Null causal effect: $\theta = 0$}                                       \\
\hline
Standard, no intercept \footnote{This is the standard inverse-variance weighted (IVW) method, equivalent to the two-stage least squares (2SLS) method with individual-level data.}
                                  &  0.000   & 0.044  &  0.047  &   3.9        &  -        \\
Standard, intercept  \footnote{This is the MR-Egger method.}
                                  &  0.002   & 0.125  &  0.133  &   3.8        &  -        \\
Robust, no intercept              &  0.000   & 0.046  &  0.050  &   4.5        &  1        \\
Robust, intercept                 &  0.002   & 0.130  &  0.141  &   5.7        &  4        \\
Penalized standard, no intercept  &  0.000   & 0.046  &  0.046  &   5.2        &  -        \\
Penalized standard, intercept     &  0.002   & 0.130  &  0.130  &   4.8        &  -        \\
Penalized robust, no intercept    &  0.000   & 0.047  &  0.047  &   5.9        &  2        \\
Penalized robust, intercept       &  0.001   & 0.132  &  0.134  &   7.1        &  5        \\
Simple median                     &  0.000   & 0.059  &  0.070  &   1.8        &  -        \\
Weighted median                   &  0.001   & 0.056  &  0.064  &   2.1        &  -        \\
Penalized weighted median         &  0.001   & 0.059  &  0.064  &   3.1        &  -        \\
L1 penalization, $\lambda = 1$    &  0.000   & 0.055  &  0.053  &   5.7        &  -        \\
L1 penalization, $\lambda = 2$    &  0.001   & 0.049  &  0.046  &   6.4        &  -        \\
L1 penalization, $\lambda = 3$    &  0.000   & 0.045  &  0.047  &   4.2        &  -        \\
L1 penalization, cross-validation &  0.001   & 0.047  &  0.046  &   5.4        &  -        \\
L1 penalization, minimal estimate &  0.000   & 0.029  &  0.062  &   0.9        &  -        \\
L1 penalization, heterogeneity    &  0.000   & 0.045  &  0.047  &   4.0        &  -        \\
\hline
\multicolumn{6}{c}{Positive causal effect: $\theta = +0.1$}                                \\
\hline
Standard, no intercept \footnotemark[2]
                                  &  0.096   & 0.047  &  0.050  &   49.3       &  -        \\
Standard, intercept    \footnotemark[3]
                                  &  0.065   & 0.136  &  0.141  &    6.7       &  -        \\
Robust, no intercept              &  0.096   & 0.048  &  0.052  &   46.1       &  2        \\
Robust, intercept                 &  0.064   & 0.140  &  0.148  &    8.7       &  4        \\
Penalized standard, no intercept  &  0.096   & 0.049  &  0.048  &   51.8       &  -        \\
Penalized standard, intercept     &  0.064   & 0.140  &  0.137  &    8.3       &  -        \\
Penalized robust, no intercept    &  0.096   & 0.050  &  0.050  &   50.1       &  2        \\
Penalized robust, intercept       &  0.064   & 0.142  &  0.140  &   10.3       &  5        \\
Simple median                     &  0.101   & 0.063  &  0.074  &   24.7       &  -        \\
Weighted median                   &  0.093   & 0.059  &  0.067  &   25.7       &  -        \\
Penalized weighted median         &  0.093   & 0.062  &  0.067  &   26.5       &  -        \\
L1 penalization, $\lambda = 1$    &  0.097   & 0.059  &  0.056  &   40.8       &  -        \\
L1 penalization, $\lambda = 2$    &  0.097   & 0.052  &  0.048  &   51.9       &  -        \\
L1 penalization, $\lambda = 3$    &  0.096   & 0.047  &  0.049  &   50.0       &  -        \\
L1 penalization, cross-validation &  0.096   & 0.051  &  0.049  &   50.4       &  -        \\
L1 penalization, minimal estimate &  0.064   & 0.048  &  0.067  &   19.2       &  -        \\
L1 penalization, heterogeneity    &  0.096   & 0.048  &  0.049  &   48.9       &  -        \\
\hline
\end{tabular}
\caption{Mean, standard deviation (SD), mean standard error (mean SE) of estimates, and empirical power (\%) from weighted linear regression models (weights are penalized where indicated) using standard and robust regression, without and with an intercept term, and median-based methods for Scenario 1.} \label{simresults.1a}
\end{small} %
\end{center}
\end{minipage}
\end{table}
\setlength{\tabcolsep}{6pt}

\setlength{\tabcolsep}{6pt} %
\begin{table}[htbp]
\begin{minipage}{\textwidth}
\begin{center}
\begin{footnotesize}
\centering
\begin{tabular}[c]{c|ccc|ccc|ccc}
\hline
                       & \multicolumn{3}{c|}{Scenario 2}  & \multicolumn{3}{c|}{Scenario 3}  & \multicolumn{3}{c}{Scenario 4}   \\
Method                 &  Mean    & SD     &  Power       &  Mean    & SD     &  Power       &  Mean    & SD     &  Power       \\
\hline
\multicolumn{10}{c}{Null causal effect: $\theta = 0$}                                                                           \\
\hline
\multicolumn{10}{c}{Proportion of invalid instrumental variables: 0.1}                                                          \\
\hline
Standard, no intercept &  0.000   & 0.069  &   5.4        &  0.067   & 0.067  &  13.0        &  0.059   & 0.076  &  19.0        \\
Standard, intercept    &  0.001   & 0.197  &   5.6        &  0.002   & 0.192  &   5.8        &  0.148   & 0.245  &  30.7        \\
Robust, no intercept   &  0.000   & 0.052  &   5.1        &  0.022   & 0.054  &   6.1        &  0.017   & 0.058  &   5.6        \\
Robust, intercept      &  0.001   & 0.153  &   6.8        &  0.001   & 0.154  &   6.5        &  0.081   & 0.205  &  11.5        \\
Simple median          &  0.000   & 0.066  &   2.8        &  0.030   & 0.066  &   4.5        &  0.013   & 0.066  &   3.2        \\
Weighted median        &  0.000   & 0.061  &   3.1        &  0.024   & 0.061  &   4.3        &  0.040   & 0.077  &  10.6        \\
\hline
\multicolumn{10}{c}{Proportion of invalid instrumental variables: 0.2}                                                          \\
\hline
Standard, no intercept & -0.001   & 0.087  &   6.1        &  0.135   & 0.084  &  34.8        &  0.112   & 0.087  &  34.0        \\
Standard, intercept    &  0.003   & 0.251  &   6.2        &  0.008   & 0.234  &   6.1        &  0.255   & 0.260  &  43.9        \\
Robust, no intercept   &  0.000   & 0.064  &   5.4        &  0.062   & 0.074  &  10.4        &  0.049   & 0.080  &   9.7        \\
Robust, intercept      &  0.003   & 0.192  &   7.1        &  0.006   & 0.192  &   7.0        &  0.207   & 0.269  &  24.9        \\
Simple median          & -0.001   & 0.074  &   3.8        &  0.067   & 0.078  &  11.3        &  0.027   & 0.074  &   4.9        \\
Weighted median        &  0.000   & 0.069  &   4.5        &  0.054   & 0.072  &  11.5        &  0.088   & 0.103  &  26.4        \\
\hline
\multicolumn{10}{c}{Proportion of invalid instrumental variables: 0.3}                                                          \\
\hline
Standard, no intercept &  0.001   & 0.103  &   5.9        &  0.204   & 0.093  &  59.3        &  0.161   & 0.092  &  48.9        \\
Standard, intercept    &  0.000   & 0.288  &   5.9        &  0.005   & 0.263  &   6.0        &  0.327   & 0.256  &  50.6        \\
Robust, no intercept   &  0.001   & 0.082  &   5.7        &  0.122   & 0.100  &  19.7        &  0.099   & 0.102  &  19.3        \\
Robust, intercept      &  0.003   & 0.243  &   6.3        &  0.005   & 0.239  &   7.5        &  0.332   & 0.293  &  42.0        \\
Simple median          &  0.000   & 0.082  &   5.1        &  0.115   & 0.094  &  25.0        &  0.046   & 0.084  &   8.6        \\
Weighted median        &  0.001   & 0.079  &   6.7        &  0.094   & 0.090  &  24.3        &  0.148   & 0.125  &  46.2        \\
\hline
\hline
\multicolumn{10}{c}{Positive causal effect: $\theta = +0.1$}                                                                    \\
\hline
\multicolumn{10}{c}{Proportion of invalid instrumental variables: 0.1}                                                          \\
\hline
Standard, no intercept &  0.095   & 0.070  &  32.6        &  0.162   & 0.069  &  69.2        &  0.155   & 0.078  &  63.3        \\
Standard, intercept    &  0.066   & 0.202  &   7.1        &  0.066   & 0.196  &   7.0        &  0.221   & 0.252  &  38.0        \\
Robust, no intercept   &  0.095   & 0.055  &  40.6        &  0.120   & 0.057  &  53.8        &  0.114   & 0.062  &  45.2        \\
Robust, intercept      &  0.066   & 0.162  &   9.0        &  0.066   & 0.162  &   8.8        &  0.149   & 0.218  &  15.8        \\
Simple median          &  0.100   & 0.070  &  23.4        &  0.132   & 0.070  &  38.5        &  0.114   & 0.070  &  29.4        \\
Weighted median        &  0.093   & 0.064  &  24.8        &  0.117   & 0.064  &  37.1        &  0.134   & 0.081  &  45.3        \\
\hline
\multicolumn{10}{c}{Proportion of invalid instrumental variables: 0.2}                                                          \\
\hline
Standard, no intercept &  0.095   & 0.089  &  22.6        &  0.230   & 0.085  &  84.0        &  0.208   & 0.089  &  73.2        \\
Standard, intercept    &  0.068   & 0.255  &   7.3        &  0.073   & 0.238  &   7.0        &  0.335   & 0.266  &  52.0        \\
Robust, no intercept   &  0.096   & 0.067  &  32.5        &  0.162   & 0.078  &  58.5        &  0.148   & 0.084  &  45.6        \\
Robust, intercept      &  0.069   & 0.201  &   9.1        &  0.071   & 0.201  &   8.8        &  0.278   & 0.281  &  29.5        \\
Simple median          &  0.100   & 0.077  &  22.9        &  0.170   & 0.083  &  53.3        &  0.129   & 0.078  &  33.9        \\
Weighted median        &  0.093   & 0.072  &  24.6        &  0.149   & 0.077  &  50.8        &  0.185   & 0.107  &  62.5        \\
\hline
\multicolumn{10}{c}{Proportion of invalid instrumental variables: 0.3}                                                          \\
\hline
Standard, no intercept &  0.097   & 0.105  &  18.6        &  0.299   & 0.095  &  93.8        &  0.257   & 0.094  &  81.6        \\
Standard, intercept    &  0.065   & 0.291  &   6.5        &  0.070   & 0.267  &   6.9        &  0.411   & 0.261  &  59.9        \\
Robust, no intercept   &  0.096   & 0.085  &  25.2        &  0.223   & 0.102  &  64.2        &  0.200   & 0.104  &  52.7        \\
Robust, intercept      &  0.068   & 0.251  &   7.7        &  0.070   & 0.247  &   8.8        &  0.406   & 0.304  &  46.7        \\
Simple median          &  0.101   & 0.087  &  22.5        &  0.221   & 0.100  &  69.2        &  0.148   & 0.089  &  38.9        \\
Weighted median        &  0.094   & 0.083  &  24.9        &  0.191   & 0.095  &  65.8        &  0.245   & 0.128  &  76.8        \\
\hline
\end{tabular}
\caption{Mean, standard deviation (SD) of estimates, and empirical power (\%) from weighted linear regression models (weights are not penalized) using standard and robust regression, without and with an intercept term, and simple and weighted median methods for Scenarios 2, 3, and 4. (Note: power with a null causal effect is the Type 1 error rate.)} \label{simresults.2a}
\end{footnotesize} %
\end{center}
\end{minipage}
\end{table}
\setlength{\tabcolsep}{6pt}

\setlength{\tabcolsep}{4pt} 
\begin{table}[htbp]
\begin{minipage}{\textwidth}
\begin{center}
\begin{footnotesize}
\centering
\begin{tabular}[c]{c|ccc|ccc|ccc}
\hline
                                 & \multicolumn{3}{c|}{Scenario 2}  & \multicolumn{3}{c|}{Scenario 3}  & \multicolumn{3}{c}{Scenario 4}   \\ 
Method                           &  Mean    & SD     &  Power       &  Mean    & SD     &  Power       &  Mean    & SD     &  Power       \\
\hline
\multicolumn{10}{c}{Null causal effect: $\theta = 0$}                                                                                     \\
\hline
\multicolumn{10}{c}{Proportion of invalid instrumental variables: 0.1}                                                                    \\
\hline
Penalized standard, no intercept &  0.000   & 0.051  &   6.8        &  0.022   & 0.053  &   9.5        &  0.019   & 0.056  &  11.3        \\
Penalized standard, intercept    &  0.001   & 0.149  &   7.5        &  0.001   & 0.154  &   8.2        &  0.093   & 0.198  &  22.7        \\
Penalized robust, no intercept   &  0.000   & 0.052  &   6.3        &  0.018   & 0.053  &   8.4        &  0.015   & 0.055  &   8.2        \\
Penalized robust, intercept      &  0.001   & 0.151  &   7.9        &  0.001   & 0.153  &   7.9        &  0.077   & 0.191  &  15.4        \\
Penalized weighted median        &  0.001   & 0.063  &   3.7        &  0.011   & 0.063  &   4.0        &  0.016   & 0.074  &   6.6        \\
\hline
\multicolumn{10}{c}{Proportion of invalid instrumental variables: 0.2}                                                                    \\
\hline
Penalized standard, no intercept &  0.000   & 0.059  &   9.6        &  0.056   & 0.068  &  24.6        &  0.049   & 0.074  &  25.2        \\
Penalized standard, intercept    &  0.004   & 0.178  &  10.9        &  0.006   & 0.194  &  13.5        &  0.206   & 0.241  &  43.4        \\
Penalized robust, no intercept   &  0.000   & 0.059  &   7.0        &  0.046   & 0.066  &  15.8        &  0.038   & 0.070  &  13.2        \\
Penalized robust, intercept      &  0.003   & 0.178  &   9.1        &  0.005   & 0.190  &  11.4        &  0.177   & 0.241  &  28.5        \\
Penalized weighted median        &  0.000   & 0.070  &   5.3        &  0.026   & 0.073  &   7.2        &  0.050   & 0.108  &  16.8        \\
\hline
\multicolumn{10}{c}{Proportion of invalid instrumental variables: 0.3}                                                                    \\
\hline
Penalized standard, no intercept &  0.001   & 0.069  &  13.3        &  0.106   & 0.088  &  46.9        &  0.091   & 0.092  &  43.9        \\
Penalized standard, intercept    &  0.003   & 0.210  &  14.6        &  0.005   & 0.247  &  20.7        &  0.309   & 0.259  &  61.4        \\
Penalized robust, no intercept   &  0.001   & 0.068  &   7.4        &  0.087   & 0.085  &  27.8        &  0.070   & 0.089  &  20.7        \\
Penalized robust, intercept      &  0.004   & 0.213  &   9.4        &  0.005   & 0.238  &  14.3        &  0.281   & 0.270  &  43.1        \\
Penalized weighted median        &  0.001   & 0.079  &   6.4        &  0.051   & 0.092  &  12.9        &  0.104   & 0.144  &  31.7        \\
\hline
\hline
\multicolumn{10}{c}{Positive causal effect: $\theta = +0.1$}                                                                              \\
\hline
\multicolumn{10}{c}{Proportion of invalid instrumental variables: 0.1}                                                                    \\
\hline
Penalized standard, no intercept &  0.095   & 0.054  &  49.0        &  0.119   & 0.056  &  65.2        &  0.116   & 0.060  &  61.3        \\
Penalized standard, intercept    &  0.065   & 0.158  &   9.9        &  0.066   & 0.162  &  10.5        &  0.164   & 0.209  &  29.5        \\
Penalized robust, no intercept   &  0.095   & 0.055  &  43.8        &  0.115   & 0.056  &  57.2        &  0.111   & 0.059  &  53.2        \\
Penalized robust, intercept      &  0.065   & 0.159  &  10.5        &  0.066   & 0.161  &  10.8        &  0.147   & 0.202  &  22.0        \\
Penalized weighted median        &  0.093   & 0.066  &  25.8        &  0.104   & 0.067  &  31.1        &  0.110   & 0.079  &  33.6        \\
\hline
\multicolumn{10}{c}{Proportion of invalid instrumental variables: 0.2}                                                                    \\
\hline
Penalized standard, no intercept &  0.095   & 0.062  &  46.8        &  0.154   & 0.071  &  78.4        &  0.147   & 0.077  &  72.1        \\
Penalized standard, intercept    &  0.069   & 0.186  &  13.6        &  0.070   & 0.202  &  15.3        &  0.282   & 0.252  &  51.8        \\
Penalized robust, no intercept   &  0.096   & 0.062  &  37.9        &  0.145   & 0.070  &  65.7        &  0.136   & 0.074  &  57.0        \\
Penalized robust, intercept      &  0.069   & 0.187  &  11.4        &  0.070   & 0.198  &  13.8        &  0.254   & 0.253  &  36.1        \\
Penalized weighted median        &  0.093   & 0.074  &  25.3        &  0.148   & 0.078  &  37.4        &  0.147   & 0.113  &  45.0        \\
\hline
\multicolumn{10}{c}{Proportion of invalid instrumental variables: 0.3}                                                                    \\
\hline
Penalized standard, no intercept &  0.097   & 0.073  &  46.1        &  0.205   & 0.091  &  89.3        &  0.190   & 0.096  &  82.6        \\
Penalized standard, intercept    &  0.069   & 0.219  &  16.9        &  0.069   & 0.254  &  21.6        &  0.390   & 0.270  &  68.8        \\
Penalized robust, no intercept   &  0.096   & 0.072  &  33.1        &  0.187   & 0.089  &  72.8        &  0.171   & 0.094  &  60.1        \\
Penalized robust, intercept      &  0.069   & 0.221  &  12.0        &  0.070   & 0.247  &  16.6        &  0.363   & 0.282  &  50.6        \\
Penalized weighted median        &  0.093   & 0.083  &  25.1        &  0.148   & 0.097  &  46.4        &  0.203   & 0.119  &  58.4        \\
\hline
\end{tabular}
\caption{Mean, standard deviation (SD) of estimates, and empirical power (\%) from weighted linear regression models (weights are penalized) using standard and robust regression, without and with an intercept term, and penalized weighted median method for Scenarios 2, 3, and 4.} \label{simresults.2b}
\end{footnotesize} %
\end{center}
\end{minipage}
\end{table}
\setlength{\tabcolsep}{6pt}

\setlength{\tabcolsep}{4pt} 
\begin{table}[htbp]
\begin{minipage}{\textwidth}
\begin{center}
\begin{footnotesize}
\centering
\begin{tabular}[c]{c|ccc|ccc|ccc}
\hline
                                 & \multicolumn{3}{c|}{Scenario 2}  & \multicolumn{3}{c|}{Scenario 3}  & \multicolumn{3}{c}{Scenario 4}   \\ 
Method                           &  Mean    & SD     &  Power       &  Mean    & SD     &  Power       &  Mean    & SD     &  Power       \\
\hline
\multicolumn{10}{c}{Null causal effect: $\theta = 0$}                                                                             \\
\hline
\multicolumn{10}{c}{Proportion of invalid instrumental variables: 0.1}                                                            \\
\hline
$\lambda = 1$                    &  0.000   &  0.062   &   6.9   &   0.025   &  0.062   &   8.7   &   0.040   &  0.074   &  14.0  \\
$\lambda = 2$                    &  0.000   &  0.053   &   7.4   &   0.015   &  0.055   &   8.8   &   0.025   &  0.066   &  12.5  \\
$\lambda = 3$                    &  0.000   &  0.050   &   5.2   &   0.012   &  0.052   &   6.4   &   0.022   &  0.071   &  11.5  \\
Cross-validation                 &  0.001   &  0.090   &   6.0   &   0.070   &  0.103   &  10.9   &   0.118   &  0.155   &  36.0  \\
Minimal estimate                 &  0.000   &  0.028   &   1.0   &   0.012   &  0.030   &   1.7   &   0.017   &  0.042   &   3.5  \\
Heterogeneity criterion          &  0.000   &  0.062   &   5.4   &   0.028   &  0.058   &   6.2   &   0.051   &  0.092   &  20.9  \\
\hline
\multicolumn{10}{c}{Proportion of invalid instrumental variables: 0.2}                                                            \\
\hline
$\lambda = 1$                    &  0.000   &  0.069   &   8.5   &   0.057   &  0.076   &  17.2   &   0.104   &  0.119   &  31.6  \\
$\lambda = 2$                    &  0.000   &  0.059   &   8.4   &   0.037   &  0.067   &  15.2   &   0.082   &  0.122   &  29.7  \\
$\lambda = 3$                    &  0.000   &  0.058   &   7.0   &   0.034   &  0.066   &  11.8   &   0.088   &  0.136   &  31.4  \\
Cross-validation                 &  0.001   &  0.142   &   8.3   &   0.192   &  0.153   &  33.3   &   0.273   &  0.178   &  70.1  \\
Minimal estimate                 &  0.000   &  0.030   &   1.0   &   0.026   &  0.041   &   4.2   &   0.057   &  0.092   &  15.0  \\
Heterogeneity criterion          &  0.000   &  0.067   &   7.0   &   0.046   &  0.065   &  11.0   &   0.099   &  0.127   &  35.8  \\
\hline
\multicolumn{10}{c}{Proportion of invalid instrumental variables: 0.3}                                                            \\
\hline
$\lambda = 1$                    & -0.001   &  0.079   &  10.1   &   0.104   &  0.108   &  30.4   &   0.206   &  0.176   &  55.6  \\
$\lambda = 2$                    &  0.000   &  0.068   &  10.3   &   0.080   &  0.102   &  28.5   &   0.187   &  0.185   &  55.5  \\
$\lambda = 3$                    &  0.000   &  0.068   &   8.0   &   0.078   &  0.103   &  24.5   &   0.203   &  0.189   &  58.9  \\
Cross-validation                 &  0.001   &  0.189   &  10.4   &   0.321   &  0.190   &  59.9   &   0.389   &  0.161   &  88.1  \\
Minimal estimate                 &  0.000   &  0.034   &   1.4   &   0.053   &  0.072   &  11.1   &   0.136   &  0.151   &  37.2  \\
Heterogeneity criterion          &  0.000   &  0.074   &   7.7   &   0.076   &  0.091   &  19.7   &   0.184   &  0.173   &  56.5  \\
\hline
\hline
\multicolumn{10}{c}{Positive causal effect: $\theta = +0.1$}                                                                      \\
\hline
\multicolumn{10}{c}{Proportion of invalid instrumental variables: 0.1}                                                            \\
\hline
$\lambda = 1$                    &  0.095   &  0.065   &  37.5   &   0.121   &  0.066   &  53.2   &   0.138   &  0.079   &  59.4  \\
$\lambda = 2$                    &  0.095   &  0.056   &  46.8   &   0.111   &  0.058   &  57.6   &   0.123   &  0.071   &  61.9  \\
$\lambda = 3$                    &  0.095   &  0.053   &  45.0   &   0.109   &  0.055   &  54.4   &   0.121   &  0.078   &  57.0  \\
Cross-validation                 &  0.095   &  0.090   &  35.5   &   0.167   &  0.105   &  59.3   &   0.213   &  0.157   &  69.6  \\
Minimal estimate                 &  0.053   &  0.051   &  12.7   &   0.082   &  0.053   &  28.5   &   0.091   &  0.063   &  34.4  \\
Heterogeneity criterion          &  0.096   &  0.066   &  39.3   &   0.125   &  0.062   &  54.7   &   0.149   &  0.096   &  60.6  \\
\hline
\multicolumn{10}{c}{Proportion of invalid instrumental variables: 0.2}                                                            \\
\hline
$\lambda = 1$                    &  0.095   &  0.073   &  36.2   &   0.156   &  0.080   &  64.5   &   0.206   &  0.125   &  75.9  \\
$\lambda = 2$                    &  0.095   &  0.063   &  44.2   &   0.136   &  0.072   &  66.2   &   0.183   &  0.127   &  74.5  \\
$\lambda = 3$                    &  0.095   &  0.061   &  41.3   &   0.133   &  0.071   &  61.8   &   0.192   &  0.142   &  70.5  \\
Cross-validation                 &  0.094   &  0.142   &  23.6   &   0.287   &  0.156   &  75.8   &   0.370   &  0.179   &  87.6  \\
Minimal estimate                 &  0.045   &  0.054   &   9.9   &   0.104   &  0.063   &  39.8   &   0.142   &  0.109   &  52.3  \\
Heterogeneity criterion          &  0.095   &  0.071   &  35.7   &   0.144   &  0.070   &  60.2   &   0.201   &  0.132   &  71.9  \\
\hline
\multicolumn{10}{c}{Proportion of invalid instrumental variables: 0.3}                                                            \\
\hline
$\lambda = 1$                    &  0.095   &  0.083   &  34.1   &   0.206   &  0.113   &  75.5   &   0.310   &  0.179   &  86.4  \\
$\lambda = 2$                    &  0.095   &  0.072   &  41.4   &   0.180   &  0.108   &  75.5   &   0.293   &  0.188   &  86.5  \\
$\lambda = 3$                    &  0.096   &  0.071   &  38.3   &   0.180   &  0.109   &  71.4   &   0.309   &  0.192   &  85.0  \\
Cross-validation                 &  0.095   &  0.190   &  19.2   &   0.419   &  0.192   &  88.8   &   0.486   &  0.162   &  95.7  \\
Minimal estimate                 &  0.042   &  0.057   &   8.6   &   0.140   &  0.089   &  53.5   &   0.231   &  0.163   &  70.8  \\
Heterogeneity criterion          &  0.096   &  0.079   &  33.8   &   0.176   &  0.098   &  69.2   &   0.288   &  0.176   &  83.2  \\
\hline
\end{tabular}
\caption{Mean, standard deviation (SD) of estimates, and empirical power (\%) from L1 penalization methods using different strategies for choosing the tuning parameter $\lambda$ for Scenarios 2, 3, and 4.} \label{simresults.2c}
\end{footnotesize} %
\end{center}
\end{minipage}
\end{table}
\setlength{\tabcolsep}{6pt}

\setlength{\tabcolsep}{4pt}
\begin{table}[htbp]
\begin{minipage}{\textwidth}
\begin{center}
\begin{small}
\centering
\begin{tabular}[c]{c|cccc|cccc}
\hline
Proportion  & \multicolumn{4}{c|}{Simple median and robust IVW methods} & \multicolumn{4}{c}{MR-Egger intercept test} \\
invalid     & Scenario 1 &  2     &  3     &  4                         & Scenario 1 &  2     &  3     &  4           \\
\hline
\multicolumn{9}{c}{Null causal effect: $\theta = 0$}                                                                  \\
\hline
 0\%        &  0.8\%     &  -     &  -     &  -                         &  3.8\%     &  -     &  -     &  -           \\
10\%        &  -         &  1.2\% &  2.2\% &  1.5\%                     &  -         &  5.5\% &  6.4\% & 24.4\%       \\
20\%        &  -         &  1.6\% &  5.5\% &  2.5\%                     &  -         &  6.0\% &  9.4\% & 31.2\%       \\
30\%        &  -         &  2.0\% & 14.1\% &  6.2\%                     &  -         &  5.9\% & 13.1\% & 32.8\%       \\
\hline
\multicolumn{9}{c}{Positive causal effect: $\theta = +0.1$}                                                           \\
\hline
 0\%        & 21.2\%     &  -     &  -     &  -                         &  4.7\%     &  -     &  -     &  -           \\
10\%        &  -         & 19.1\% & 33.0\% & 23.6\%                     &  -         &  5.8\% &  8.0\% & 21.8\%       \\
20\%        &  -         & 16.7\% & 44.1\% & 26.6\%                     &  -         &  6.1\% & 11.1\% & 28.2\%       \\
30\%        &  -         & 14.6\% & 55.8\% & 31.8\%                     &  -         &  6.0\% & 15.3\% & 30.0\%       \\
\hline
\end{tabular}
\caption{Proportion of simulated datasets for which the simple median and robust regression with no intercept (robust IVW) methods rejected the causal null (left), empirical power of the intercept test in MR-Egger method for detecting directional pleiotropy and/or violation of the InSIDE assumption in all scenarios.} \label{simresults.3}
\end{small} %
\end{center}
\end{minipage}
\end{table}
\setlength{\tabcolsep}{6pt}

\begin{table}[htbp] 
\begin{center}
\begin{small}
\centering
\begin{tabular}[c]{c|cccc}
\hline
                                  & \multicolumn{2}{c}{Non-penalized weights}  & \multicolumn{2}{c}{Penalized weights}   \\
Method                            &  Estimate (SE)  &  95\% CI                 &  Estimate (SE)  &  95\% CI              \\
\hline
Standard, no intercept            & -0.031 (0.100)  & -0.227,  0.165           & -0.034 (0.057)  & -0.147,  0.078        \\
Standard, intercept               &  0.336 (0.241)  & -0.136,  0.808           &  0.154 (0.143)  & -0.127,  0.435        \\
Robust, no intercept              & -0.024 (0.079)  & -0.180,  0.132           & -0.033 (0.062)  & -0.154,  0.089        \\
Robust, intercept                 &  0.255 (0.212)  & -0.162,  0.671           &  0.142 (0.150)  & -0.152,  0.436        \\
Simple median                     & -0.073 (0.088)  & -0.244,  0.098           &  -              &  -                    \\
Weighted median                   & -0.075 (0.087)  & -0.246,  0.096           & -0.076 (0.090)  & -0.253,  0.100        \\
L1 pen, cross-validation          & -0.036 (0.087)  & -0.207,  0.136           &  -              &  -                    \\
L1 pen, heterogeneity             & -0.022 (0.055)  & -0.131,  0.086           &  -              &  -                    \\
\hline
\end{tabular}
\caption{Estimates (standard errors, SE) and 95\% confidence intervals (CI, calculated as estimate $\pm$ 1.96 standard errors) of causal effect of body mass index on schizophrenia risk (log odds ratio for schizophrenia per 1 standard deviation increase in body mass index).} \label{exresults}
\end{small} 
\end{center}
\end{table}
\setlength{\tabcolsep}{6pt}

\begin{table}[htbp]
\begin{minipage}{\textwidth}
\begin{adjustwidth}{-0.6cm}{-0.6cm}
\begin{center}
\begin{small}
\centering
\begin{tabular}[c]{p{0.3\textwidth}p{0.7\textwidth}}
\hline
Method                  &  Description                                                                                                                                  \\
\hline
Inverse-variance weighted (IVW) method
                        &  Standard weighted regression with inverse-variance weights and intercept term set to zero.                                                   \\ [2mm]
MR-Egger method         &  Standard weighted regression with inverse-variance weights and intercept term estimated.                                                     \\ [2mm]
Median-based method     &  Simple median method is the median of the causal estimates based on the individual candidate instruments.
             Weighted median method uses inverse-variance weights so that more precise estimates receive more weight in the analysis.                                   \\ [2mm]
Robust \mbox{regression} (MM-estimation with bisquare objective function)
                        &  Standard regression in either the IVW (no intercept) or the MR-Egger (intercept) method can be replaced with robust regression.              \\ [6mm]
Penalization of weights & Inverse-variance weights in either the IVW, MR-Egger, or weighted median method can be replaced with weights that depend on the heterogeneity
            of the causal estimates -- candidate instruments with outlying estimates are downweighted depending on the degree of heterogeneity.                         \\ [2mm]
L1 penalization         & A separate intercept term, representing the pleiotropic effect of the candidate instrument on the outcome,
            is allowed for each candidate instrument, but the sum of the absolute values of the pleiotropic effects is not allowed to be too large. The value of the
            tuning parameter, which regulates the extent to which pleiotropic effects are penalized, must be chosen carefully.                                          \\
\hline
\end{tabular}
\caption{Summary of methods investigated in this paper.} \label{summary}
\end{small} %
\end{center}
\end{adjustwidth}
\end{minipage}
\end{table}

\clearpage
\subsection*{Figures}

\clearpage
\renewcommand{\thesection}{A.\arabic{section}}
\renewcommand{\thesubsection}{A.\arabic{section}.\arabic{subsection}}
\renewcommand{\thetable}{A\arabic{table}}
\renewcommand{\thefigure}{A\arabic{figure}}
\setcounter{table}{0}
\setcounter{figure}{0}
\renewcommand{\tablename}{Web Table}
\renewcommand{\figurename}{Web Figure}
\setcounter{section}{0}
\setcounter{subsection}{0}
\section*{Web Appendix}

\section{Software code}
We provide R code to implement the methods discussed in this paper. The associations of the candidate instruments with the exposure are denoted \texttt{betaXG} with standard errors \texttt{sebetaXG}. The associations of the candidate instruments with the outcome are denoted \texttt{betaYG} with standard errors \texttt{sebetaYG}. We assume that the candidate instruments are uncorrelated in their distributions, as is common in applied Mendelian randomization investigations.

\vspace{3mm}

\subsubsection*{Inverse-variance weighted estimate:}

\normalsize{
The inverse-variance weighted (IVW) estimate can be calculated by weighted linear regression:}\par
\scriptsize{
\begin{verbatim}
betaIVW               = summary(lm(betaYG~betaXG-1, weights=sebetaYG^-2))$coef[1]
sebetaIVW.fixed       = summary(lm(betaYG~betaXG-1, weights=sebetaYG^-2))$coef[1,2]/
                         summary(lm(betaYG~betaXG-1, weights=sebetaYG^-2))$sigma
sebetaIVW.random      = summary(lm(betaYG~betaXG-1, weights=sebetaYG^-2))$coef[1,2]/
                     min(summary(lm(betaYG~betaXG-1, weights=sebetaYG^-2))$sigma,1)
\end{verbatim}
}\par

\normalsize{
In the fixed-effect model, we divide the reported standard error by the estimated residual standard error, to fix the residual standard error to take the value 1 \cite{thompson1999}. In the multiplicative random-effects model, we divide by the estimated residual standard error in the case of underdispersion (the variability in the genetic associations is less than would be expected by chance alone). But in the case of overdispersion (that is, heterogeneity of causal effect estimates), no correction is made. The point estimate is unaffected by the choice of a fixed- or multiplicative random-effects model.
}\par

\vspace{3mm}

\normalsize{
Alternatively, the inverse-variance weighted estimate can be calculated by meta-analysis, or via a simple formula:}\par
\scriptsize{
\begin{verbatim}
# meta-analysis
library(meta)
betaIVW               = metagen(betaYG/betaXG, abs(sebetaYG/betaXG))$TE.fixed
sebetaIVW.fixed       = metagen(betaYG/betaXG, abs(sebetaYG/betaXG))$seTE.fixed
# simple formula
betaIVW               = sum(betaYG*betaXG*sebetaYG^-2)/sum(betaXG^2*sebetaYG^-2)
sebetaIVW.fixed       = 1/sqrt(sum(betaXG^2*sebetaYG^-2))
\end{verbatim}
}\par

\normalsize{
The meta-analysis method can be used to perform an additive random-effects analysis, which makes a different parametric assumption about the heterogeneity between causal estimates compared with the multiplicative random-effect analysis \cite{burgess2016ivw}. While the causal estimates from the fixed-effect and multiplicative random-effects analyses are the same, the estimate from the additive random-effects analysis differs.}\par

\vspace{3mm}

\subsubsection*{MR-Egger regression:}
\normalsize{
The MR-Egger method is equivalent to the IVW method calculated using weighted regression, except that that intercept term is estimated rather than being set to zero. A test as to whether the intercept term is equal to zero is a test of directional pleiotropy. A random-effects model should be used for inference as a fixed-effect model is not justifiable when the candidate instruments are not all valid. }\par

\scriptsize{
\begin{verbatim}
# coding of genetic variants
betaYG = betaYG*sign(betaXG); betaXG = abs(betaXG)
# causal estimate
betaEGGER          = summary(lm(betaYG~betaXG, weights=sebetaYG^-2))$coef[2,1]
sebetaEGGER.random = summary(lm(betaYG~betaXG, weights=sebetaYG^-2))$coef[2,2]/
                      min(summary(lm(betaYG~betaXG, weights=sebetaYG^-2))$sigma, 1)
betaEGGER.lower    = betaEGGER-qt(0.975,df=length(betaXG)-2)*sebetaEGGER.random
betaEGGER.upper    = betaEGGER+qt(0.975,df=length(betaXG)-2)*sebetaEGGER.random
p.causal.random    = 2*(1-pt(abs(betaEGGER/sebetaEGGER.random),df=length(betaXG)-2))
# test for directional pleiotropy
interEGGER          = summary(lm(betaYG~betaXG, weights=sebetaYG^-2))$coef[1,1]
seinterEGGER.random = summary(lm(betaYG~betaXG, weights=sebetaYG^-2))$coef[1,2]/
                   min(summary(lm(betaYG~betaXG, weights=sebetaYG^-2))$sigma, 1)
p.dpleio.random     = 2*(1-pt(abs(interEGGER/seinterEGGER.random),df=length(betaXG)-2))
\end{verbatim}
}\par

\normalsize{
In this code, we use a t-distribution with $J-2$ degrees of freedom for inference. If there is underdispersion, then the t-distribution may be overly conservative, as the t-distribution assumes that the residual standard error is estimated (in case of underdispersion, the residual standard error is set to 1). Hence, if the residual standard error is less than one, either a confidence interval using a residual standard error of 1 and a z-distribution, or else a confidence interval using the estimated residual standard error and a t-distribution may be preferred (the wider of these two intervals should be preferred -- both of these will be narrower than the above confidence interval). }\par

\scriptsize{
\begin{verbatim}
sigmaEGGER         = summary(lm(betaYG~betaXG, weights=sebetaYG^-2))$sigma
betaEGGER.lower    = ifelse(sigmaEGGER<1, min(betaEGGER-qnorm(0.975)*sebetaEGGER.random,
                  betaEGGER-qt(0.975,df=length(betaXG)-2)*sebetaEGGER.random*sigmaEGGER),
                  betaEGGER-qt(0.975,df=length(betaXG)-2)*sebetaEGGER.random)
betaEGGER.upper    = ifelse(sigmaEGGER<1, max(betaEGGER+qnorm(0.975)*sebetaEGGER.random,
                  betaEGGER+qt(0.975,df=length(betaXG)-2)*sebetaEGGER.random*sigmaEGGER),
                  betaEGGER+qt(0.975,df=length(betaXG)-2)*sebetaEGGER.random)
\end{verbatim}
}\par

\normalsize{\hphantom{v}}\par

\vspace{3mm}

\subsubsection*{Median-based method:}
\normalsize{
The median-based method calculates the median (or weighted median) of the causal estimates from each candidate instrument. This code calculates the simple median, weighted median, and penalized weighted median, employing bootstrapping to obtain a standard error that can used to provide a confidence interval. }\par

\scriptsize{
\begin{verbatim}
weighted.median <- function(betaIV.in, weights.in) {
  betaIV.order  = betaIV.in[order(betaIV.in)]
  weights.order = weights.in[order(betaIV.in)]
  weights.sum   = cumsum(weights.order)-0.5*weights.order
  weights.sum   = weights.sum/sum(weights.order)
        below   = max(which(weights.sum<0.5))
  weighted.est  = betaIV.order[below] + (betaIV.order[below+1]-betaIV.order[below])*
                  (0.5-weights.sum[below])/(weights.sum[below+1]-weights.sum[below])
  return(weighted.est) }
 #
weighted.median.boot = function(betaXG.in, betaYG.in, sebetaXG.in, sebetaYG.in, weights.in){
 # the standard error is estimated based on 1000 bootstrap samples
med = NULL
for(i in 1:1000){
 betaXG.boot = rnorm(length(betaXG.in), mean=betaXG.in, sd=sebetaXG.in)
 betaYG.boot = rnorm(length(betaYG.in), mean=betaYG.in, sd=sebetaYG.in)
 betaIV.boot  = betaYG.boot/betaXG.boot
 med[i] = weighted.median(betaIV.boot, weights.in)
 }
return(sd(med)) }
 #
betaIV             = betaYG/betaXG
weights            = rep(1, length(betaXG)) # unweighted median
betaSIMPLEMED      = weighted.median(betaIV, weights)
sebetaSIMPLEMED    = weighted.median.boot(betaXG, betaYG, sebetaXG, sebetaYG, weights)
lowerSIMPLEMED     = betaSIMPLEMED-qnorm(0.975)*sebetaSIMPLEMED
upperSIMPLEMED     = betaSIMPLEMED+qnorm(0.975)*sebetaSIMPLEMED
 #
betaIV             = betaYG/betaXG
weights            = (sebetaYG/betaXG)^-2   # weighted median using inverse-variance weights
betaWEIGHTEDMED    = weighted.median(betaIV, weights)
sebetaWEIGHTEDMED  = weighted.median.boot(betaXG, betaYG, sebetaXG, sebetaYG, weights)
lowerWEIGHTEDMED   = betaWEIGHTEDMED-qnorm(0.975)*sebetaWEIGHTEDMED
upperWEIGHTEDMED   = betaWEIGHTEDMED+qnorm(0.975)*sebetaWEIGHTEDMED
 #
betaIV             = betaYG/betaXG          # penalized weighted median
penalty            = pchisq(weights*(betaIV-betaWEIGHTEDMED)^2, df=1, lower.tail=FALSE)
pen.weights        = (sebetaYG/betaXG)^-2*pmin(1, penalty*20)    # penalized weights
betaPENALIZEDMED   = weighted.median(betaIV, pen.weights)
sebetaPENALIZEDMED = weighted.median.boot(betaXG, betaYG, sebetaXG, sebetaYG, pen.weights)
lowerPENALIZEDMED  = betaPENALIZEDMED-qnorm(0.975)*sebetaPENALIZEDMED
upperPENALIZEDMED  = betaPENALIZEDMED+qnorm(0.975)*sebetaPENALIZEDMED
\end{verbatim}
}\par

\normalsize{\hphantom{v}}\par

\subsubsection*{Robust regression:}
\normalsize{
The IVW and MR-Egger methods can be performed using robust regression (in particular, MM-estimation using Tukey's bisquare objective function) rather than standard linear regression:}\par
\scriptsize{
\begin{verbatim}
library(robustbase)
betaIVW.robust            = summary(lmrob(betaYG~betaXG-1, weights=sebetaYG^-2, k.max=500))$coef[1]
sebetaIVW.robust.fixed    = summary(lmrob(betaYG~betaXG-1, weights=sebetaYG^-2, k.max=500))$coef[1,2]/
                            summary(lmrob(betaYG~betaXG-1, weights=sebetaYG^-2, k.max=500))$sigma
sebetaIVW.robust.random   = summary(lmrob(betaYG~betaXG-1, weights=sebetaYG^-2, k.max=500))$coef[1,2]/
                        min(summary(lmrob(betaYG~betaXG-1, weights=sebetaYG^-2, k.max=500))$sigma,1)
betaEGGER.robust          = summary(lmrob(betaYG~betaXG,   weights=sebetaYG^-2, k.max=500))$coef[2]
sebetaEGGER.robust.random = summary(lmrob(betaYG~betaXG,   weights=sebetaYG^-2, k.max=500))$coef[2,2]/
                        min(summary(lmrob(betaYG~betaXG,   weights=sebetaYG^-2, k.max=500))$sigma,1)
\end{verbatim}
}\par
\normalsize{The \texttt{k.max} option sets the maximum number of steps evaluated to find initial parameter values in the S-step of the algorithm.}\par

\subsubsection*{Penalized weights:}
\normalsize{
The IVW and MR-Egger methods can be performed using penalized weights:}\par
\scriptsize{
\begin{verbatim}
betaIVW            = sum(betaYG*betaXG*sebetaYG^-2)/sum(betaXG^2*sebetaYG^-2)
pweights  = pchisq(betaXG^2/sebetaYG^2*(betaYG/betaXG-betaIVW)^2, df=1, lower.tail=FALSE)
pweightsE = pchisq(sebetaYG^-2*(betaYG - interEGGER - betaEGGER*betaXG)^2, df=1, lower.tail=FALSE)
rweights  = sebetaYG^-2*pmin(1, pweights*20)
rweightsE = sebetaYG^-2*pmin(1, pweightsE*20)
betaIVW.penal            = summary(lm(betaYG~betaXG-1, weights=rweights))$coef[1]
sebetaIVW.penal.fixed    = summary(lm(betaYG~betaXG-1, weights=rweights))$coef[1,2]/
                           summary(lm(betaYG~betaXG-1, weights=rweights))$sigma
sebetaIVW.penal.random   = summary(lm(betaYG~betaXG-1, weights=rweights))$coef[1,2]/
                       min(summary(lm(betaYG~betaXG-1, weights=rweights))$sigma,1)
betaEGGER.penal          = summary(lm(betaYG~betaXG,   weights=rweightsE))$coef[2]
sebetaEGGER.penal.random = summary(lm(betaYG~betaXG,   weights=rweightsE))$coef[2,2]/
                       min(summary(lm(betaYG~betaXG,   weights=rweightsE))$sigma,1)
\end{verbatim}
}\par
\normalsize{Penalized weights can also be used in conjunction with robust regression.}\par

\subsubsection*{L1 penalization:}
\normalsize{
Several packages are available for running various flavours of L1 penalization methods. We chose the \emph{penalized} package as this gave an option for some of the coefficients in the model to be penalized (the pleiotropy intercept parameters), and others not to be penalized (the causal effect parameter):}\par
\footnotesize{
\begin{verbatim}
library(penalized)
betaYGw = betaYG/sebetaYG  # dividing the association estimates by sebetaYG is equivalent
betaXGw = betaXG/sebetaYG  # to weighting by sebetaYG^-2
pleio = diag(rep(1, length(betaXG)))

l1one_which = which(attributes(penalized(betaYGw, pleio, betaXGw, lambda1=1))$penalized==0)
l1one_beta  = lm(betaYG[l1one_which]~betaXG[l1one_which]-1, weights=sebetaYG[l1one_which]^-2)$coef[1]
l1one_se    = summary(lm(betaYG[l1one_which]~betaXG[l1one_which]-1, weights=sebetaYG[l1one_which]^-2))$coef[1,2]
         /min(summary(lm(betaYG[l1one_which]~betaXG[l1one_which]-1, weights=sebetaYG[l1one_which]^-2))$sigma, 1)

l1two_which = which(attributes(penalized(betaYGw, pleio, betaXGw, lambda1=2))$penalized==0)
l1two_beta  = lm(betaYG[l1two_which]~betaXG[l1two_which]-1, weights=sebetaYG[l1two_which]^-2)$coef[1]
l1two_se    = summary(lm(betaYG[l1two_which]~betaXG[l1two_which]-1, weights=sebetaYG[l1two_which]^-2))$coef[1,2]
         /min(summary(lm(betaYG[l1two_which]~betaXG[l1two_which]-1, weights=sebetaYG[l1two_which]^-2))$sigma, 1)

l1three_which = which(attributes(penalized(betaYGw, pleio, betaXGw, lambda1=3))$penalized==0)
l1three_beta  = lm(betaYG[l1three_which]~betaXG[l1three_which]-1, weights=sebetaYG[l1three_which]^-2)$coef[1]
l1three_se    = summary(lm(betaYG[l1three_which]~betaXG[l1three_which]-1, weights=sebetaYG[l1three_which]^-2))$coef[1,2]
           /min(summary(lm(betaYG[l1three_which]~betaXG[l1three_which]-1, weights=sebetaYG[l1three_which]^-2))$sigma, 1)
 # fixing lambda to be 1, 2, and 3 in turn

l1grid = c(seq(from=0.1, to=5, by=0.1), seq(from=5.2, to=10, by=0.2))
    # values of lambda for grid search
l1grid_rse = NULL; l1grid_length = NULL; l1grid_beta = NULL; l1grid_se = NULL

for (i in 1:length(l1grid)) {
 l1grid_which = which(attributes(penalized(betaYGw, pleio, betaXGw, lambda1=l1grid[i], trace=FALSE))$penalized==0)
 l1grid_rse[i]    = summary(lm(betaYG[l1grid_which]~betaXG[l1grid_which]-1, weights=sebetaYG[l1grid_which]^-2))$sigma
 l1grid_length[i] = length(l1grid_which)
 l1grid_beta[i]   = lm(betaYG[l1grid_which]~betaXG[l1grid_which]-1, weights=sebetaYG[l1grid_which]^-2)$coef[1]
 l1grid_se[i]     = summary(lm(betaYG[l1grid_which]~betaXG[l1grid_which]-1, weights=sebetaYG[l1grid_which]^-2))$coef[1,2]/
                min(summary(lm(betaYG[l1grid_which]~betaXG[l1grid_which]-1, weights=sebetaYG[l1grid_which]^-2))$sigma, 1)
 }

l1which_hetero = c(which(l1grid_rse[1:(length(l1grid)-1)]>1&
                    diff(l1grid_rse)>qchisq(0.95, df=1)/l1grid_length[2:length(l1grid)]), length(l1grid))[1]
    # heterogeneity criterion for choosing lambda
l1hetero_beta  = l1grid_beta[l1which_hetero]
l1hetero_se    = l1grid_se[l1which_hetero]

l1which_min    = which.min(l1grid_beta)
l1min_beta     = l1grid_beta[l1which_min]
l1min_se       = l1grid_se[l1which_min]
    # minimal estimate criterion for choosing lambda

l1xval_lambda = optL1(betaYGw, pleio, betaXGw)$lambda
l1xval_which  = which(attributes(penalized(betaYGw, pleio, betaXGw, lambda1=l1xval_lambda))$penalized==0)
l1xval_beta   = summary(lm(alpy[l1xval_which]~alpx[l1xval_which]-1, weights=alpysd[l1xval_which]^-2))$coef[1]
l1xval_se     = summary(lm(alpy[l1xval_which]~alpx[l1xval_which]-1, weights=alpysd[l1xval_which]^-2))$coef[1,2]/
            min(summary(lm(alpy[l1xval_which]~alpx[l1xval_which]-1, weights=alpysd[l1xval_which]^-2))$sigma, 1)
    # cross-validation criterion for choosing lambda
\end{verbatim}
}\par
\normalsize{We found that the choice of values of $\lambda$ for the grid search presented here ($0.1, 0.2, \ldots, 4.9, 5.0$, $5.2, 5.4, \ldots, 9.8, 10.0$) worked well in both the simulations and the applied example. However, for different sets of association estimates, a different choice of values may be preferred. Additionally, particularly with large numbers of variants, a more dense choice of values may be preferred to ensure that at must one variant is added to the analysis at each incremental step.}

\clearpage

\section{Choice of penalty function}
When there are two candidate instruments, the use of an L1 penalty function is equivalent to minimizing:
\begin{equation}
S_1 = (Y_1 - \theta_{01} - \theta_{1} X_1)^2 + (Y_2 - \theta_{01} - \theta_{1} X_2)^2 + 2 \lambda (|\theta_{01}| +  |\theta_{02}|) \notag
\end{equation}
where $Y_j$ is $\hat{\beta}_{Yj}/\se(\hat{\beta}_{Yj})$ and $X_j$ is $\hat{\beta}_{Xj}/\se(\hat{\beta}_{Yj})$. The factor of two on the penalty function and the change of notation are for simplicity of presentation, and dividing the associations by $\se(\hat{\beta}_{Yj})$ is equivalent to inverse-variance weighting.

Differentiating this expression, we get:
\begin{align}
\frac{\partial S_1}{\partial \theta_{01}} &= -2 (Y_1 - \theta_{01} - \theta_1 X_1) + 2 \lambda \sign(\theta_{01}) \notag \\
\frac{\partial S_1}{\partial \theta_{02}} &= -2 (Y_2 - \theta_{02} - \theta_1 X_2) + 2 \lambda \sign(\theta_{02}) \notag \\
\frac{\partial S_1}{\partial \theta_1}    &= -2 X_1 (Y_1 - \theta_{01} - \theta_1 X_1) - 2 X_2 (Y_2 - \theta_{02} - \theta X_2) \notag
\end{align}

As $S_1$ is not continuous, this function is minimized either at a discontinuity ($\theta_{01} = 0$, $\theta_{02} = 0$), or where the derivatives equal zero. If $\hat{\theta}_{01}$ and $\hat{\theta}_{02}$ both differ from zero, then:
\begin{align}
\hat{\theta}_{01} &= Y_1 - \hat{\theta}_1 X_1 - \sign(\hat{\theta}_{01}) \lambda \notag \\
\hat{\theta}_{02} &= Y_2 - \hat{\theta}_1 X_1 - \sign(\hat{\theta}_{02}) \lambda \notag \\
\hat{\theta}_1    &= \frac{-\lambda (X_1 \sign(\hat{\theta}_{01}) + X_2 \sign(\hat{\theta}_{02}))}{2 (X_1^2 + X_2^2)} \notag
\end{align}

When $\lambda$ is close to zero, $\hat{\theta}_{01}$ and $\hat{\theta}_{02}$ will both differ from zero, whereas when $\lambda$ is large, $\hat{\theta}_{01}$ and $\hat{\theta}_{02}$ will both equal zero. The upshot is that the value of $\hat{\theta}_1$ depends on the value of $\lambda$.

In contrast, if we were to use an L2 penalty function, we would minimize:
\begin{equation}
S_2 = (Y_1 - \theta_{01} - \theta_{1} X_1)^2 + (Y_2 - \theta_{01} - \theta_{1} X_2)^2 + \lambda (\theta_{01}^2 +  \theta_{02}^2) \notag
\end{equation}

This function is continuous, and so its minimum is where the partial derivatives equal zero:
\begin{align}
\frac{\partial S_2}{\partial \theta_{01}} &= -2 (Y_1 - \theta_{01} - \theta_1 X_1) + 2 \lambda \theta_{01} \notag \\
\frac{\partial S_2}{\partial \theta_{02}} &= -2 (Y_2 - \theta_{02} - \theta_1 X_2) + 2 \lambda \theta_{02} \notag \\
\frac{\partial S_2}{\partial \theta_1}    &= -2 X_1 (Y_1 - \theta_{01} - \theta_1 X_1) - 2 X_2 (Y_2 - \theta_{02} - \theta X_2) \notag \\
\hat{\theta}_{01} &= \frac{Y_1 - \hat{\theta}_1 X_1}{1 + \lambda} \notag \\
\hat{\theta}_{02} &= \frac{Y_2 - \hat{\theta}_1 X_1}{1 + \lambda} \notag \\
\hat{\theta}_1    &= \frac{X_1 Y_1 + X_2 Y_2}{X_1^2 + X_2^2} \notag
\end{align}

The causal estimate is not a function of $\lambda$. Hence, L2 penalization cannot be used either for robust estimation, or for identifying valid instruments (as it does not have a sparsity property; $\hat{\theta}_{01}$ and $\hat{\theta}_{02}$ differ from zero for all finite values of $\lambda$).

\clearpage

\section{Supplementary tables for simulation study}
\subsection{Number of simulations that failed to report a standard error}
The numbers of simulations for the robust methods that failed to report a standard error in Scenarios 2 to 4 are provided in Web Table~\ref{simresults.na}. The proportion of simulations was usually less than 1\%, and was less than 2.5\% in all cases.

\begin{table}[htbp]
\begin{minipage}{\textwidth}
\begin{center}
\begin{small}
\centering
\begin{tabular}[c]{c|ccc|ccc}
\hline
Method                           & \multicolumn{3}{c|}{Null causal effect}               & \multicolumn{3}{c}{Positive causal effect}            \\
\hline
\multicolumn{1}{r|}{Proportion invalid:}
                                 & 10\%         & 20\%         & 30\%                    & 10\%         &  20\%         & 30\%                   \\
\hline
\multicolumn{7}{c}{Scenario 2: balanced pleiotropy, InSIDE satisfied}                                                                            \\
\hline
Robust, no intercept             &   1          &   2          &   0                     &   0          &   1          &   0                     \\
Robust, intercept                &   5          &  20          &  24                     &  12          &  18          &  24                     \\
Penalized robust, no intercept   &   4          &   5          &  18                     &   0          &  10          &  12                     \\
Penalized robust, intercept      &  15          &  42          &  97                     &  10          &  23          &  80                     \\
\hline
\multicolumn{7}{c}{Scenario 3: directional pleiotropy, InSIDE satisfied}                                                                         \\
\hline
Robust, no intercept             &   1          &   0          &   0                     &   0          &   1          &   1                     \\
Robust, intercept                &   2          &  18          &  15                     &   5          &  10          &  11                     \\
Penalized robust, no intercept   &   4          &   5          &   8                     &   1          &   2          &   3                     \\
Penalized robust, intercept      &   2          &  30          &  44                     &  15          &  19          &  31                     \\
\hline
\multicolumn{7}{c}{Scenario 4: directional pleiotropy, InSIDE not satisfied}                                                                     \\
\hline
Robust, no intercept             &   2          &  15          &  34                     &   3          &  15          &  26                     \\
Robust, intercept                & 131          & 245          & 244                     & 147          & 233          & 227                     \\
Penalized robust, no intercept   &   2          &  12          &  44                     &   2          &   9          &  41                     \\
Penalized robust, intercept      &  24          &  69          & 102                     &  31          &  51          &  92                     \\
\hline
\end{tabular}
\caption{Number of the 10\thinspace000 simulations that failed to report a standard error using the robust regression method in each of the simulation settings.} \label{simresults.na}
\end{small} %
\end{center}
\end{minipage}
\end{table}

\subsection{One-sample setting}
The simulation study from the main body of the paper was repeated, except in a one-sample setting in which associations of the candidate instruments with the exposure and with the outcome were obtained in the same sample of 20\thinspace000 individuals for the methods using non-penalized weights. Results are displayed in Web Table~\ref{simresults.1one} (Scenario 1) and Web Table~\ref{simresults.2one} (Scenarios 2 to 4).

\begin{table}[htbp] 
\begin{minipage}{\textwidth}
\begin{center}
\begin{small}
\centering
\begin{tabular}[c]{c|ccccc}
\hline
                                 & \multicolumn{3}{c}{Scenario 1}                         \\
Method                           &  Mean    & SD     &  Mean SE  &  Power       &  NA
   \footnote{Number of the 10\thinspace000 simulations that failed to report a standard error.} \\
\hline
\multicolumn{6}{c}{Null causal effect: $\theta = 0$}                                      \\
\hline
Standard, no intercept \footnote{This is the standard inverse-variance weighted (IVW) method.}
                                 &  0.024   & 0.044  &  0.047  &   6.8        &  -        \\
Standard, intercept  \footnote{This is the MR-Egger method.}
                                 &  0.173   & 0.123  &  0.131  &  27.2        &  -        \\
Robust, no intercept             &  0.023   & 0.044  &  0.049  &   7.8        &  0        \\
Robust, intercept                &  0.174   & 0.126  &  0.136  &  29.0        &  4        \\
Penalized standard, no intercept &  0.024   & 0.045  &  0.046  &   8.5        &  -        \\
Penalized standard, intercept    &  0.174   & 0.126  &  0.129  &  28.5        &  -        \\
Penalized robust, no intercept   &  0.024   & 0.046  &  0.047  &   9.4        &  0        \\
Penalized robust, intercept      &  0.174   & 0.127  &  0.131  &  31.0        &  3        \\
Simple median                    &  0.000   & 0.060  &  0.070  &   2.0        &  -        \\
Weighted median                  &  0.038   & 0.054  &  0.063  &   5.5        &  -        \\
Penalized weighted median        &  0.038   & 0.057  &  0.063  &   6.5        &  -        \\
\hline
\multicolumn{6}{c}{Positive causal effect: $\theta = +0.1$}                               \\
\hline
Standard, no intercept \footnotemark[2]
                                 &  0.123   & 0.044  &  0.048  &   73.6       &  -        \\
Standard, intercept    \footnotemark[3]
                                 &  0.271   & 0.121  &  0.136  &   53.1       &  -        \\
Robust, no intercept             &  0.122   & 0.045  &  0.050  &   69.3       &  0        \\
Robust, intercept                &  0.271   & 0.125  &  0.143  &   52.4       &  3        \\
Penalized standard, no intercept &  0.122   & 0.045  &  0.048  &   74.1       &  -        \\
Penalized standard, intercept    &  0.271   & 0.123  &  0.135  &   53.8       &  -        \\
Penalized robust, no intercept   &  0.122   & 0.046  &  0.049  &   71.6       &  1        \\
Penalized robust, intercept      &  0.271   & 0.126  &  0.139  &   53.9       &  2        \\
Simple median                    &  0.099   & 0.059  &  0.073  &   25.4       &  -        \\
Weighted median                  &  0.136   & 0.054  &  0.067  &   54.6       &  -        \\
Penalized weighted median        &  0.136   & 0.057  &  0.067  &   54.5       &  -        \\
\hline
\end{tabular}
\caption{Mean, standard deviation (SD), mean standard error (mean SE) of estimates, and empirical power (\%) from weighted linear regression models (weights are penalized where indicated) using standard and robust regression, without and with an intercept term, and median-based methods for Scenario 1 in one-sample setting (associations with exposure and with outcome are estimated in the same individuals).} \label{simresults.1one}
\end{small} %
\end{center}
\end{minipage}
\end{table}
\setlength{\tabcolsep}{6pt}

\setlength{\tabcolsep}{6pt} %
\begin{table}[htbp]
\begin{minipage}{\textwidth}
\begin{center}
\begin{footnotesize}
\centering
\begin{tabular}[c]{c|ccc|ccc|ccc}
\hline
                       & \multicolumn{3}{c|}{Scenario 2}  & \multicolumn{3}{c|}{Scenario 3}  & \multicolumn{3}{c}{Scenario 4}   \\
Method                 &  Mean    & SD     &  Power       &  Mean    & SD     &  Power       &  Mean    & SD     &  Power       \\
\hline
\multicolumn{10}{c}{Null causal effect: $\theta = 0$}                                                                           \\
\hline
\multicolumn{10}{c}{Proportion of invalid instrumental variables: 0.1}                                                          \\
\hline
Standard, no intercept &  0.023   & 0.069  &   7.3        &  0.090   & 0.066  &  24.5        &  0.081   & 0.072  &  27.1        \\
Standard, intercept    &  0.174   & 0.197  &  19.4        &  0.175   & 0.190  &  19.2        &  0.276   & 0.228  &  50.7        \\
Robust, no intercept   &  0.023   & 0.052  &   8.3        &  0.045   & 0.053  &  13.2        &  0.039   & 0.057  &  10.4        \\
Robust, intercept      &  0.174   & 0.148  &  26.9        &  0.175   & 0.148  &  27.1        &  0.263   & 0.187  &  36.9        \\
Simple median          & -0.001   & 0.066  &   2.9        &  0.028   & 0.064  &   4.5        &  0.012   & 0.065  &   3.2        \\
Weighted median        &  0.037   & 0.061  &   6.8        &  0.061   & 0.060  &  12.6        &  0.074   & 0.074  &  20.0        \\
\hline
\multicolumn{10}{c}{Proportion of invalid instrumental variables: 0.2}                                                          \\
\hline
Standard, no intercept &  0.022   & 0.087  &   6.8        &  0.158   & 0.082  &  50.4        &  0.133   & 0.084  &  44.3        \\
Standard, intercept    &  0.172   & 0.248  &  13.9        &  0.178   & 0.232  &  14.9        &  0.354   & 0.243  &  58.8        \\
Robust, no intercept   &  0.023   & 0.063  &   7.7        &  0.084   & 0.072  &  19.4        &  0.072   & 0.078  &  15.5        \\
Robust, intercept      &  0.172   & 0.187  &  23.0        &  0.175   & 0.186  &  22.6        &  0.382   & 0.225  &  51.9        \\
Simple median          & -0.001   & 0.073  &   3.5        &  0.064   & 0.073  &  11.2        &  0.026   & 0.071  &   4.7        \\
Weighted median        &  0.037   & 0.067  &   7.8        &  0.089   & 0.069  &  22.5        &  0.121   & 0.096  &  39.0        \\
\hline
\multicolumn{10}{c}{Proportion of invalid instrumental variables: 0.3}                                                          \\
\hline
Standard, no intercept &  0.024   & 0.102  &   6.8        &  0.226   & 0.091  &  74.7        &  0.181   & 0.088  &  58.8        \\
Standard, intercept    &  0.171   & 0.286  &  11.0        &  0.178   & 0.259  &  12.8        &  0.404   & 0.243  &  61.8        \\
Robust, no intercept   &  0.022   & 0.080  &   6.7        &  0.143   & 0.096  &  30.3        &  0.120   & 0.098  &  26.8        \\
Robust, intercept      &  0.172   & 0.234  &  17.5        &  0.176   & 0.232  &  17.6        &  0.485   & 0.239  &  65.5        \\
Simple median          & -0.001   & 0.082  &   4.7        &  0.109   & 0.085  &  24.7        &  0.043   & 0.080  &   8.0        \\
Weighted median        &  0.036   & 0.077  &   9.2        &  0.127   & 0.084  &  38.4        &  0.176   & 0.117  &  58.7        \\
\hline
\hline
\multicolumn{10}{c}{Positive causal effect: $\theta = +0.1$}                                                                    \\
\hline
\multicolumn{10}{c}{Proportion of invalid instrumental variables: 0.1}                                                          \\
\hline
Standard, no intercept &  0.123   & 0.069  &  49.4        &  0.190   & 0.066  &  87.2        &  0.181   & 0.072  &  79.3        \\
Standard, intercept    &  0.274   & 0.197  &  35.7        &  0.275   & 0.190  &  36.4        &  0.376   & 0.228  &  65.1        \\
Robust, no intercept   &  0.123   & 0.052  &  62.8        &  0.145   & 0.053  &  74.8        &  0.139   & 0.057  &  67.3        \\
Robust, intercept      &  0.274   & 0.148  &  49.2        &  0.275   & 0.148  &  49.2        &  0.363   & 0.187  &  55.7        \\
Simple median          &  0.099   & 0.066  &  24.6        &  0.128   & 0.064  &  38.9        &  0.112   & 0.065  &  30.2        \\
Weighted median        &  0.137   & 0.061  &  52.0        &  0.161   & 0.060  &  66.0        &  0.174   & 0.074  &  70.1        \\
\hline
\multicolumn{10}{c}{Proportion of invalid instrumental variables: 0.2}                                                          \\
\hline
Standard, no intercept &  0.123   & 0.087  &  34.9        &  0.258   & 0.082  &  95.2        &  0.233   & 0.084  &  84.8        \\
Standard, intercept    &  0.272   & 0.248  &  25.2        &  0.278   & 0.232  &  27.2        &  0.454   & 0.243  &  69.9        \\
Robust, no intercept   &  0.123   & 0.063  &  52.4        &  0.184   & 0.072  &  76.3        &  0.172   & 0.078  &  64.8        \\
Robust, intercept      &  0.273   & 0.187  &  40.7        &  0.275   & 0.186  &  40.5        &  0.482   & 0.225  &  64.3        \\
Simple median          &  0.099   & 0.073  &  23.9        &  0.164   & 0.073  &  54.6        &  0.126   & 0.071  &  35.1        \\
Weighted median        &  0.137   & 0.067  &  49.8        &  0.189   & 0.069  &  76.4        &  0.221   & 0.096  &  81.4        \\
\hline
\multicolumn{10}{c}{Proportion of invalid instrumental variables: 0.3}                                                          \\
\hline
Standard, no intercept &  0.124   & 0.102  &  26.8        &  0.326   & 0.091  &  98.1        &  0.281   & 0.088  &  89.3        \\
Standard, intercept    &  0.271   & 0.286  &  19.0        &  0.278   & 0.259  &  21.4        &  0.504   & 0.243  &  73.3        \\
Robust, no intercept   &  0.122   & 0.080  &  40.2        &  0.243   & 0.096  &  76.8        &  0.220   & 0.098  &  66.7        \\
Robust, intercept      &  0.272   & 0.234  &  30.6        &  0.276   & 0.232  &  31.0        &  0.585   & 0.239  &  74.3        \\
Simple median          &  0.099   & 0.082  &  23.2        &  0.209   & 0.085  &  69.5        &  0.143   & 0.080  &  40.8        \\
Weighted median        &  0.136   & 0.077  &  46.6        &  0.227   & 0.084  &  84.6        &  0.276   & 0.117  &  89.1        \\
\hline
\end{tabular}
\caption{Mean, standard deviation (SD), mean standard error (mean SE) of estimates, and empirical power (\%) from weighted linear regression models (weights are not penalized) using standard and robust regression, without and with an intercept term, and simple and weighted median methods for Scenarios 2, 3, and 4 in one-sample setting (associations with exposure and with outcome are estimated in the same individuals).} \label{simresults.2one}
\end{footnotesize} %
\end{center}
\end{minipage}
\end{table}
\setlength{\tabcolsep}{6pt}

\subsection{Fewer candidate instruments}
The simulation was also repeated in a two-sample setting with only 10 candidate instruments, to observe whether the robust methods were able to operate well with fewer instruments to detect violations of the instrumental variables assumptions. Results for Scenarios 2 to 4 are presented in Web Table~\ref{simresults.2ten}.

\setlength{\tabcolsep}{6pt} %
\begin{table}[htbp]
\begin{minipage}{\textwidth}
\begin{center}
\begin{footnotesize}
\centering
\begin{tabular}[c]{c|ccc|ccc|ccc}
\hline
                       & \multicolumn{3}{c|}{Scenario 2}  & \multicolumn{3}{c|}{Scenario 3}  & \multicolumn{3}{c}{Scenario 4}   \\
Method                 &  Mean    & SD     &  Power       &  Mean    & SD     &  Power       &  Mean    & SD     &  Power       \\
\hline
\multicolumn{10}{c}{Null causal effect: $\theta = 0$}                                                                           \\
\hline
\multicolumn{10}{c}{Proportion of invalid instrumental variables: 0.1}                                                          \\
\hline
Standard, no intercept & -0.001   & 0.113  &   5.5        & 0.068   & 0.109  &    7.2        & 0.053   & 0.112  &   11.4        \\
Standard, intercept    & -0.003   & 0.353  &   6.2        & 0.001   & 0.343  &    6.3        & 0.123   & 0.366  &   20.0        \\
Robust, no intercept   &  0.000   & 0.090  &   7.0        & 0.030   & 0.093  &    7.4        & 0.024   & 0.098  &    8.4        \\
Robust, intercept      &  0.001   & 0.321  &  12.5        & 0.002   & 0.315  &   12.3        & 0.092   & 0.348  &   19.7        \\
Simple median          & -0.001   & 0.105  &   3.2        & 0.033   & 0.106  &    4.0        & 0.013   & 0.103  &    3.0        \\
Weighted median        &  0.000   & 0.098  &   4.2        & 0.028   & 0.100  &    4.9        & 0.044   & 0.124  &   11.0        \\
\hline
\multicolumn{10}{c}{Proportion of invalid instrumental variables: 0.2}                                                          \\
\hline
Standard, no intercept &  0.001   & 0.139  &   5.8        & 0.136   & 0.133  &   13.9        & 0.101   & 0.133  &   18.4        \\
Standard, intercept    &  0.001   & 0.432  &   7.1        & 0.002   & 0.412  &    7.5        & 0.213   & 0.418  &   30.8        \\
Robust, no intercept   &  0.002   & 0.111  &   7.2        & 0.076   & 0.126  &    9.9        & 0.058   & 0.125  &   12.6        \\
Robust, intercept      &  0.002   & 0.398  &  14.3        & 0.001   & 0.387  &   14.2        & 0.185   & 0.418  &   29.5        \\
Simple median          &  0.000   & 0.117  &   4.0        & 0.075   & 0.129  &    8.2        & 0.028   & 0.117  &    5.0        \\
Weighted median        &  0.001   & 0.113  &   5.8        & 0.063   & 0.125  &    9.9        & 0.092   & 0.152  &   21.8        \\
\hline
\multicolumn{10}{c}{Proportion of invalid instrumental variables: 0.3}                                                          \\
\hline
Standard, no intercept &  0.001   & 0.166  &   7.0        & 0.205   & 0.151  &   24.7        & 0.148   & 0.145  &   25.3        \\
Standard, intercept    &  0.011   & 0.510  &   7.8        & 0.008   & 0.461  &    7.8        & 0.287   & 0.426  &   35.7        \\
Robust, no intercept   &  0.000   & 0.140  &   7.8        & 0.140   & 0.160  &   15.9        & 0.102   & 0.148  &   20.4        \\
Robust, intercept      &  0.002   & 0.494  &  13.4        & 0.003   & 0.455  &   14.7        & 0.271   & 0.453  &   36.7        \\
Simple median          &  0.001   & 0.140  &   6.3        & 0.134   & 0.164  &   17.6        & 0.052   & 0.136  &    8.6        \\
Weighted median        &  0.001   & 0.136  &   8.6        & 0.110   & 0.156  &   17.7        & 0.144   & 0.173  &   34.5        \\
\hline
\hline
\multicolumn{10}{c}{Positive causal effect: $\theta = +0.1$}                                                                    \\
\hline
\multicolumn{10}{c}{Proportion of invalid instrumental variables: 0.1}                                                          \\
\hline
Standard, no intercept &  0.095   & 0.116  &  18.1        &  0.164   & 0.112  &  30.3        &  0.149   & 0.116  &  32.1        \\
Standard, intercept    &  0.063   & 0.361  &   6.8        &  0.067   & 0.352  &   6.7        &  0.196   & 0.376  &  22.8        \\
Robust, no intercept   &  0.096   & 0.095  &  20.8        &  0.128   & 0.098  &  26.1        &  0.122   & 0.103  &  25.2        \\
Robust, intercept      &  0.066   & 0.337  &  13.3        &  0.068   & 0.328  &  13.7        &  0.160   & 0.363  &  21.8        \\
Simple median          &  0.101   & 0.111  &  11.8        &  0.136   & 0.113  &  17.9        &  0.115   & 0.109  &  14.2        \\
Weighted median        &  0.093   & 0.104  &  13.0        &  0.122   & 0.106  &  18.8        &  0.140   & 0.130  &  25.3        \\
\hline
\multicolumn{10}{c}{Proportion of invalid instrumental variables: 0.2}                                                          \\
\hline
Standard, no intercept &  0.097   & 0.141  &  15.4        &  0.232   & 0.136  &  40.9        &  0.197   & 0.136  &  39.7        \\
Standard, intercept    &  0.067   & 0.439  &   7.4        &  0.068   & 0.420  &   8.1        &  0.292   & 0.427  &  34.5        \\
Robust, no intercept   &  0.097   & 0.116  &  18.7        &  0.175   & 0.130  &  30.3        &  0.157   & 0.129  &  30.1        \\
Robust, intercept      &  0.068   & 0.411  &  14.9        &  0.067   & 0.401  &  15.1        &  0.259   & 0.432  &  31.8        \\
Simple median          &  0.101   & 0.123  &  12.1        &  0.180   & 0.136  &  26.7        &  0.130   & 0.124  &  17.3        \\
Weighted median        &  0.094   & 0.118  &  14.0        &  0.159   & 0.130  &  26.7        &  0.188   & 0.158  &  37.7        \\
\hline
\multicolumn{10}{c}{Proportion of invalid instrumental variables: 0.3}                                                          \\
\hline
Standard, no intercept &  0.097   & 0.167  &  14.4        &  0.301   & 0.154  &  53.0        &  0.244   & 0.147  &  46.9        \\
Standard, intercept    &  0.077   & 0.516  &  8.3         &  0.074   & 0.467  &   8.0        &  0.371   & 0.434  &  39.9        \\
Robust, no intercept   &  0.096   & 0.144  &  16.9        &  0.240   & 0.162  &  38.4        &  0.202   & 0.151  &  37.0        \\
Robust, intercept      &  0.068   & 0.505  &  14.4        &  0.068   & 0.464  &  15.5        &  0.349   & 0.465  &  39.4        \\
Simple median          &  0.102   & 0.146  &  14.1        &  0.240   & 0.171  &  39.9        &  0.155   & 0.142  &  23.4        \\
Weighted median        &  0.094   & 0.142  &  16.4        &  0.207   & 0.160  &  38.1        &  0.242   & 0.178  &  51.2        \\
\hline
\end{tabular}
\caption{Mean, standard deviation (SD), mean standard error (mean SE) of estimates, and empirical power (\%) from weighted linear regression models (weights are not penalized) using standard and robust regression, without and with an intercept term, and simple and weighted median methods for Scenarios 2, 3, and 4 in two-sample setting with only 10 candidate instruments (25 candidate instruments are used in all other simulations).} \label{simresults.2ten}
\end{footnotesize} %
\end{center}
\end{minipage}
\end{table}
\setlength{\tabcolsep}{6pt}

\end{document}